\RequirePackage{fix-cm}
\documentclass[12pt,a4paper,final]{amsart}

\usepackage{graphicx}
\usepackage[export]{adjustbox}
\usepackage{mathptmx}      % use Times fonts if available on your TeX system
\usepackage{amssymb}% http://ctan.org/pkg/amssymb
\usepackage{pifont}% http://ctan.org/pkg/pifont
\usepackage{subfiles}
\usepackage{rotating}
\usepackage{pdflscape}
\usepackage{wasysym}
\usepackage{bookmark}
\usepackage{tablefootnote}
\usepackage{xcolor}
\usepackage{url}
\makeatletter
\def\Url@twoslashes{\mathchar`\/\@ifnextchar/{\kern-.2em}{}}
\g@addto@macro\UrlSpecials{\do\/{\Url@twoslashes}}
\makeatother
\usepackage{hyperref}

\newcommand{\alphabin}{\alpha_{binary}}
\newcommand{\cualpha}{cu\textrm{-}\alpha}
\newcommand{\Cualpha}{Cu\textrm{-}\alpha}

\newcommand{\cC}{\mathcal{C}}
\newcommand{\cS}{\mathcal{S}}

\begin{document}

\newtheorem{thm}{Theorem}[section]
\theoremstyle{remark}
\newtheorem{remark}[thm]{Remark}

\title{Why are many businesses instilling a DevOps culture\\into their organization?}


\author[]{Jessica Diaz}
\address{ETSI de Sistemas Inform\'aticos, Universidad Polit\'ecnica de Madrid. C.\ Alan Turing s/n, 28031. Madrid, Spain.}
\email{yesica.diaz@upm.es}
\author[]{Daniel L\'opez-Fern\'andez}
\address{ETSI de Sistemas Inform\'aticos, Universidad Polit\'ecnica de Madrid. C.\ Alan Turing s/n, 28031. Madrid, Spain.}
\email{daniel.lopez@upm.es}
\author[]{Jorge Perez}
\address{ETSI de Sistemas Inform\'aticos, Universidad Polit\'ecnica de Madrid. C.\ Alan Turing s/n, 28031. Madrid, Spain.}
\email{jorgeenrique.perez@upm.es}
\author[]{\'Angel Gonz\'alez-Prieto}
\address{ETSI de Sistemas Inform\'aticos, Universidad Polit\'ecnica de Madrid. C.\ Alan Turing s/n, 28031. Madrid, Spain.}
\email{angel.gonzalez.prieto@upm.es}


\maketitle

\begin{abstract}
\textbf{Context}: DevOps can be defined as a cultural movement to improve and accelerate the delivery of business value by making the collaboration between development and operations effective. Although this movement is relatively recent, there exist an intensive research around DevOps. However, the real reasons why companies move to DevOps and the results they expect to obtain have been paid little attention in real contexts. 
\textbf{Objective}: This paper aims to help practitioners and researchers to better understand the context and the problems that many companies face day to day in their organizations when they try to accelerate software delivery and the main drivers that move these companies to adopting DevOps.
\textbf{Method}: We conducted an exploratory study by leveraging in depth, semi-structured interviews to relevant stakeholders of 30 multinational software-intensive companies, together industrial workshops and observations at organizations' facilities that supported triangulation. Additionally, we conducted an inter-coder agreement analysis, which is not usually addressed in qualitative studies in software engineering, to increase reliability and reduce authors bias of the drawn findings. 
\textbf{Results}: The research explores the problems and expected outcomes that moved companies to adopt DevOps and reveals a set of patterns and anti-patterns about the reasons why companies are instilling a DevOps culture. \textbf{Conclusions}: This study aims to strengthen evidence and support practitioners in making better informed about which problems trigger a DevOps transition and most common expected results.


\end{abstract}

\section{Introduction}
\label{intro}
The current digital landscape requires shortening the time-to-market from the generation of business ideas until the production of the software that supports these ideas. High customer expectations force organizations to adopt an experimental organizational culture through which organizations constantly develop new business ideas and test these ideas with their customers \cite{Bosch:2012}. This is motivated, and in turn supported, by a prevalent business model based on the web and a software-as-a-service model over cloud infrastructure \cite{Bosch:2012}. Companies that approach this continuous experimentation and that can release software early and frequently have a greater capacity for business innovation, and thus a higher capability to compete in the market. Innovative companies, such as Google, Amazon, Facebook or Spotify, are characterized by short releases and quick response time to customer demands, being able to make multiple deploys per day \cite{GoogleAmazon,Facebook,Spotify}. These companies have shifted much of the software engineering landscape, mainly in terms of organizational culture and practices \cite{Parnin:2017}. The agile \cite{Beck:2001} and lean \cite{Poppendieck:2007} software development approaches are at the forefront of this shift. However, despite the considerable improvements in software development that these approaches bring \cite{Dingsoyr:2012,Rodriguez:2018}, one of the main problems that slows down speed and agility is the existence of a very strict division of responsibilities between IT departments, which become silos that produce delays in software delivery and decrease software quality. Specifically, we refer to the separation in the value chain between those who develop new product features and those who put these features into production. 

During 2008-2009, leading software companies started an organizational transformation to break down these organizational silos and to shift towards fast and frequent delivery of software updates to the customer \cite{Fitzgerald:2017}. DevOps breaks down organizational silos and \textit{``stresses empathy and cross-functional collaboration within and between teams—especially development and IT operations—in order to operate resilient systems and accelerate delivery of changes''} \cite{Dyck:2015}. DevOps requires the development team to work closely and efficiently with the operations team to enable continuous deployment, i.e., to put every change into production through the automation of deployment pipelines, resulting in many production deployments every day \cite{Fowler:2017,Leppanen:2015}. Thus, DevOps is an organizational approach—also referred to as a cultural movement and a technical solution—that aims to deploy software updates on continuous basis to the production environment while also ensuring reliable operability of the live environment \cite{Lwakatare:2016}.

Some reports \cite{dora:2018,puppet:2018} show that DevOps currently plays a fundamental role in large software-intensive organizations whose business greatly depends on how efficient development and operations are. However, this movement is relatively recent, and in addition to considering the cases of these large and well-known companies, it is necessary to collect more empirical evidence about the reasons that motivate companies to adopt DevOps and what results they expect to obtain when adopting DevOps culture.

This paper empirically investigates the practice of DevOps in various software development companies. The method is an exploratory multiple case study of 30 multinational software-intensive companies through interviews with relevant stakeholders, together with industrial workshops and observations at organizations' facilities that supported triangulation and allowed us to complete data omissions in the interviews when necessary. This study aims to help practitioners and researchers better understand the context and the problems that many companies are facing in their organizations to accelerate innovation and software delivery and the main drivers that move these companies to adopt DevOps, as well as the results they expect. The study contributes to strengthening the evidence regarding DevOps and supports practitioners in making better informed decisions in DevOps transformation processes, specifically by highlighting the problems that commonly trigger a DevOps transition and the most common expected results. There are some decisions that can lead one organization to failure and many others to success, so the only way to be sure of being on the right track is to follow decisions that have been successfully made on numerous occasions. 

Furthermore, to motivate others to provide similar evidence to help mature DevOps research and practice by replicating the study, we have made available the instruments used in this study, as well as the methods used to increase the reliability of findings, specifically inter-coder agreement analysis based on Krippendorff's coefficients.

The structure of the paper is as follows: Section~\ref{back} provides an overview of the rise of the DevOps culture. Section~\ref{method} describes the method used for the exploratory case study. Section~\ref{results} reports the results and findings and Section~\ref{validity} assesses the statistical validity of these outcomes. Section~\ref{related} describes related work. Finally, conclusions and further work are presented in Section~\ref{conclusions}.

\section{Background}
\label{back}
DevOps is an organizational transformation that had its origin at the 2008 Agile Conference in Toronto, where P. Debois highlighted the need to resolve the conflict between development and operations teams when they had to collaborate to provide quick response time to customer demands \cite{Debois:2008}. Later, at the O’Reilly Velocity Conference, two Flickr employees delivered a seminal talk known as \textit{``10+ Deploys per Day: Dev and Ops Cooperation at Flickr''}, which can be considered the starting point to extend agility beyond development \cite{Allspaw:2009}. Currently, an entire industry has been created around DevOps tools whose objective is to automatize best practices, such as continuous delivery and continuous deployment. These practices promote the fast and frequent delivery of new and changing features while ensuring the quality and non-disruption of the production environment and customers \cite{Lwakatare:2016}.

However, beyond all that, DevOps is a cultural movement that aims for collaboration among all stakeholders involved in the development, deployment and operation of software to deliver a high-quality product or service in the shortest possible time. It is a simple concept, but its adoption by organizations is enormously complicated because of great differences in the way in which DevOps promotes work and the traditional way in which most software companies have been working for decades. As that transformation requires great effort by companies, their CEOs, CIOs, and practitioners in general need evidence about the problems and drivers that are currently moving companies to adopt DevOps. According to the DevOps Agile Skills Association (DASA\footnote{\url{https://www.devopsagileskills.org/}, last accessed 2020/01/01.}), the main drivers are making IT easier, faster and cheaper, and providing more business value. This means reducing time-to-market, accelerating innovation, reducing costs, enhancing team communication and collaboration, reducing errors, and improving system stability, among others. This paper empirically studies the practice of DevOps in various software development companies to examine the problems and drivers that move companies to adopt DevOps and the results they expect to achieve.
%Are these findings supported by empirical evidence?

%DevOps is founded on the Lean principles and shares its values, such as process optimization, search for continuous improvement, and the enhancement of customer satisfaction.

\section{Research Methodology: Study Design}
\label{method}
This paper presents empirical research on practicing DevOps. It is mainly based on the \textit{constructivism} model as an underlying philosophy \cite{Easterbrook:2008}. Constructivism or interpretivism states that scientific knowledge cannot be separated from its human context, and a phenomenon can be fully understood by considering the perspectives and the context of the involved participants. Therefore, the most suitable methods to support this approach are those collecting rich qualitative data, from which theories (tied to the context under study) may emerge. The research methodology of this empirical study on practicing DevOps is a multiple case study. Exploratory case studies are useful for determining what is happening in a phenomenon while also seeking new insights and generating ideas for new research \cite{Wohlin:2012}. The study has been conducted according to the guidelines for conducting case study research in software engineering proposed by Runeson and Höst \cite{Runeson:2009}. The collection methods we used are interviews, workshops, observations, and retrieval of a set of metrics.

The specific investigation presented here characterizes the reasons why companies move to DevOps and the results they expect to obtain when adopting DevOps culture, and it has been conducted mainly by leveraging in-depth, semi-structured interviews with software practitioners from 30 multinational software-intensive companies from November 2017 to February 2020. Interviews are a common method used for collecting data in software field studies \cite{Lethbridge:2005}. The interviews provide us with the necessary data to answer the research questions of this study. Additionally, observations at organizations' facilities and industrial workshops provided us with data and memos that we used to triangulate and complete data omissions. The results of the study are presented through an exploratory analysis.

The details of the research methodology were previously described, discussed, and improved at the \textit{Fostering More Industry-Academic Research in XP (FIAREX)} workshop, part of XP 2018 conference \cite{Diaz:2018}, and the \textit{Product-Focused Software Process Improvement} 2019 conference \cite{Diaz:2019}. We have established a chain of evidence by following a strict process. First, we prepared the script of an interview. Afterwards, interviews were performed, recorded, and transcribed. To qualitatively analyze the data, we used the thematic analysis approach, \cite{Thomas:2008,Cruzes:2011}, which is one of the most commonly used synthesis methods that consists of coding, grouping, interconnecting and obtaining patterns. Finally, we performed a detailed analysis of the results and their validity, with special emphasis on reliability and the reduction of the authors' bias.

\subsection{Research Questions}
The research questions to be answered through the analysis of this multiple case study can be formulated as follows:
\begin{itemize}
    \item \textbf{RQ1} What \textbf{problems} do companies try to solve by implementing DevOps?
    \item \textbf{RQ2} What \textbf{results} do companies try to achieve by implementing DevOps?
\end{itemize}

\subsection{Data Collections and Instruments}
The main data collection method was semi-structured interviews with software practitioners of 30 companies. The interviews were conducted face-to-face by two researchers from November 2017 to December 2019. The interviews took approximately 2.5 hours, although in some cases, we contacted the interviewees several times.

We prepared a script for the interview that was first tested with five organizations to evaluate its suitability. The questions of the interview were collected from existing survey studies on the state of DevOps \cite{dora:2018,puppet:2018,Kim:2016}, exploratory studies \cite{Erich:2017,Lwakatare:2016}, and meetings with experts in international workshops (FIAREX workshop \cite{Diaz:2018} and PROFES conference \cite{Diaz:2019}). The interview includes a set of close-ended demographic questions, open questions, and semi-open questions. The questions were refined as we gained more knowledge during the interviews and workshops. 

The interview is structured to collect professional information about the interviewees, organizations, context and problems before addressing DevOps adoption, drivers and expected results, DevOps adoption processes, organizational and team structure, culture-related practices, team-related practices, collaboration-related practices, sharing-related practices, automation-related practices, measurement and monitoring-related practices, barriers, and results. %All questions focus on a unit of analysis, i.e. the department/team in which the interviewee is/was active, and one product/service or project being developed under the DevOps culture and practices.  
The full interview script and the rest of the case study material are available through the project's web \url{https://blogs.upm.es/devopsinpractice} and repositories.  

The interviews were conducted face-to-face using the Spanish language, and the audio was recorded with the permission of the participants, transcribed for the purpose of data analysis, and reviewed by the respondents. In the transcripts, each case (i.e., the organization) was given an individual identification number, as shown in Table~\ref{tab:Subjects}. 

Furthermore, on many occasions, we had the opportunity to attend working sessions in the organizations’ facilities. In these work sessions, we observed IT departments in their daily work. In this way, we observed first-hand product poly-skilled teams, development teams and infrastructure teams, how they collaborated (e.g., they showed us some content from Confluence, Jira, Slack) or how they did not collaborate. In some cases, they even showed us information radiators, where we collected data related to problems (e.g., delays in releases, system downtime, open issues, fail rates of pipeline executions). On other occasions, we organized industrial workshops with some participant organizations at the university campus\footnote{\url{http://bit.ly/2ky00LQ}, last accessed 2020/01/01.}. Therefore, we managed to involve the participants in the study beyond the interviews. Finally, we can mention that some participants showed us different snapshots of a DevOps maturity framework, from which we could discover the problems that motivated them to adopt DevOps in early stages, how change leaders convinced their CIO with these data and the expected results, and how they started the DevOps adoption. We obtained many observations of these visits (gathered in a \textit{research diary}) that we analyzed together in interviews, mainly to triangulate data and complete data omissions.

The study was promoted through personal contacts of the participating researchers, conferences, professional associations, and networks and, to a lesser extent, posts to social media channels (Twitter and LinkedIn). 

\subsection{Subject Description}
We targeted software-intensive organizations with +2 years of experience in the adoption of DevOps. The sampling for the study can be considered a combination of \textit{maximum variation sampling} and \textit{convenience sampling}. The convenience sampling strategy is a non-probability/non-random sampling technique used to create samples as per ease of access to organizations and relevant stakeholders. Most companies participating in the study were contacted at DevOps-related events, such as DevOps Spain\footnote{\url{https://www.devops-spain.com/}  last accessed 2020/01/01.},  itSMF events\footnote{\url{http://bit.ly/2ky0eCG }, last accessed 2020/01/01.}, and DevOpsDays\footnote{\url{https://devopsdays.org/events/2020-madrid/welcome/} last accessed 2020/01/01.}, among others. This may lead to organizations not fully reflecting the entire target audience (e.g., organizations that do not attend these events), but guarantees that the organizations involved in this study are representative of the DevOps movement and its culture. This and other threats to validity are analyzed in Section~\ref{method-ica}.

%For each interview, we selected key stakeholders who are members of DevOps teams or know very well the daily work of these teams. If an interviewee did not know to answer a question, that question was answered later after the interviewee obtained that knowledge. 
Table~\ref{tab:Subjects} lists the organizations involved in the study, its ID, scope (international or national), size\footnote{Spanish Law 5/2015 indicates that a \textit{micro enterprise} is one that has less than ten workers and an annual turnover of less than two million euros or a total asset of less than two million euros; a \textit{small company} is one that has a maximum of 49 workers and a turnover or total assets of less than ten million euros;  \textit{medium-sized companies} are those with less than 250 workers and a turnover of less than fifty million euros or an asset of less than 43 million euros; and \textit{large companies} are those that exceed these parameters.}, business core and organization age. 
In many cases, more than one participant participated in an interview. Indeed, a total of 44 people participated in the 30 interviews. Table~\ref{tab:Interviewees} provides anonymized information about the position and IT experience of the interviewees. We interviewed key stakeholders, such as CEOs, CIOs, DevOps platform leaders, product leaders, developers, and infrastructure managers (see Table \ref{tab:Interviewees}). These interviewees had all the necessary information to adequately answer the questions posed. %Moreover, in some cases we contacted with the interviewees several times until all questions or some misinterpretation were solved. 

\begin{table}[t]
\centering
\caption{Subject description}
\small
\begin{tabular}{p{0.5cm} p{2cm} p{1.6cm} p{4cm} p{2.5cm}}
\hline
\textbf{Id} & \textbf{Scope} & \textbf{Size} &\textbf{Business} &\textbf{Creation date} \\ 
\hline 
01 & International & Medium & Retail &  2000-2010 \\  
02 & National & Large & Retail & $<$2000 \\ 
03 & International & Medium & Software & 2000-2010 \\
04 & National & Large & Telecom & $<$2000 \\ 
05 & National & Large & Public Utility & $<$2000 \\ 
06 & International & Large & Consulting$\Rightarrow$ Banking & $<$2000 \\ 
07 & National & Large & Educational & $<$2000  \\ 
08 & National & Large & Consulting$\Rightarrow$~N/A & $<$2000 \\ 
09 & International & Large & FinTech & 2000-2010 \\ 
10 & National & Medium & Consulting$\Rightarrow$~Logistic & $<$2000 \\ 
11 & International & Medium & Retail & $<$2000 \\ 
12 & International & Large & Logistic & $<$2000\\ 
13 & International & Large & Retail & $<$2000  \\ 
14 & International & Large & Telecom & $<$2000 \\ 
15 & National & Large & Consulting$\Rightarrow$~Telecom & $<$2000 \\ 
16 & National & Large & Consulting$\Rightarrow$~Banking & $<$2000 \\ 
17 & International & Large & Telecom & 2000-2010 \\ 
18 & International & Large & Real estate & $<$2000 \\ 
19 & International & Large & Consulting$\Rightarrow$~Banking & $<$ 2000 \\ 
20 & National & Large & Insurance & $<$2000 \\ 
21 & National & Large & Consulting$\Rightarrow$~Marketing & 2000-2010 \\ 
22 & International & Small & Consulting$\Rightarrow$~Retail & $>$2010 \\ 
23 & International & Large & Telecom & $<$2000 \\ 
24 & International & Large &  Consulting$\Rightarrow$~N/A & $<$2000 \\ 
25 & International & Large & Consulting$\Rightarrow$~Telecom  & $<$2000 \\ 
26 & National & Large & Banking & $<$ 2000 \\ 
27 & International & Large & Consulting$\Rightarrow$~N/A & $<$2000 \\ 
28 & International & Large & Marketplace & 2000-2010 \\ 
29 & International & Large & Retail & $<$2000 \\ 
30 & International & Large & Consulting$\Rightarrow$~Banking & 2010 \\ 
\hline
\end{tabular}
\label{tab:Subjects}
\end{table}

\begin{table}[t]
\centering
\caption{(Anonymized) Description of Interviewees}
\small
\begin{tabular}{p{5cm} p{1.2cm} }
\hline
\textbf{Position} & \textbf{Number}\\ 
\hline 
Executive manager & 11 \\  
Service manager & 3 \\ 
Infrastructure manager & 7 \\
Project manager & 10 \\ 
Consultant & 5 \\ 
Developer & 8 \\ 
\hline 
\textbf{Experience (years)} & \textbf{Number}\\
\hline 
$+$20 & 16\\ 
16-20 & 12\\
11-15 & 10\\ 
5-10 & 6\\ 
\hline
\end{tabular}

\label{tab:Interviewees}
\end{table}

\subsection{Data Analysis}
\label{dataanalysis}
As this work is primary qualitative research, we used the thematic analysis approach \cite{Thomas:2008,Cruzes:2011}. Thematic analysis is a method for identifying, analyzing, and reporting patterns (themes) within data that are codified segment by segment \cite{Cruzes:2011}. \textit{Codes} are defined as \textit{``descriptive labels that are applied to segments of text from each study''} \cite{Cruzes:2011}. It is convenient to highlight that \textit{coding} is more than applying codes to segments that exemplify the same theoretical or descriptive idea \cite{Floersch:2010}: \textit{``coding requires a clear sense of the context in which findings are made"} \cite{Cruzes:2011}. \textit{Themes} result from organizing and grouping similar codes into categories that share some characteristics. Themes reduce large amounts of codes into a smaller number of analytic units, and help the researcher elaborate a cognitive map. We used Atlas.ti 8 to instrument the thematic analysis of the interviews \cite{Atlas:2019,atlas:2014}. According to the use of Atlas.ti, themes are referred to as \textit{semantic domains}. The method for data analysis we followed is described in three phases:

{\setlength{\parindent}{0pt} \textbf{Phase 1. Coding}: We applied an \textit{integrated approach} for thematic analysis \cite{Cruzes:2011} that employs both a \textit{deductive approach} \cite{Miles:1994} for creating semantic domains and an \textit{inductive approach} (grounded theory) \cite{Corbin:2007} for creating codes.}

\begin{figure}[h]
\centering
\includegraphics [width=11cm]{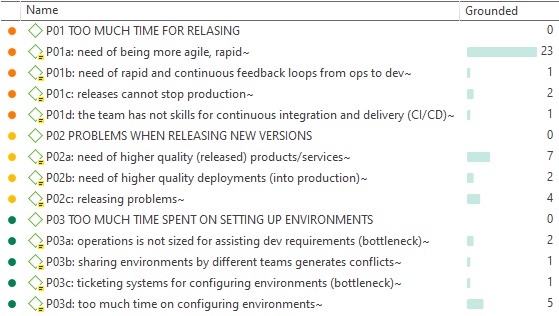}
\caption{Atlas.ti code manager}
\label{fig:1}
\end{figure}

First (deductive approach), Researcher 1 (first author) created a list of semantic domains in which codes would be later grouped inductively. These initial domains integrate concepts known in the literature and discussed in the abovementioned workshops and events. For domains related to RQ1 (problems), each domain is named P01, P02, P03, etc. For domains related to RQ2 (results), each domain is named R01, R02, R03, etc. Domains were written with uppercase letters (see Figure~\ref{fig:1}).

Second (inductive approach), Researcher 1 approached the data (i.e., the transcriptions of the interviews) with RQ1 and RQ2 in mind. Researcher 1 reviewed the data line by line, created \textit{quotations} (segments of text), and assigned them a new code or a previously defined code. As more interviews were analyzed the resulting \textit{codebook} was refined by using a \textit{constant comparison method} that forced the researcher to go back and forth. 

Additionally, the codes were commented on to explicitly define the concept they describe, in such a way that they must satisfy two requirements that Atlas.ti defines as follows \cite{Atlas:2019}: \textit{exhaustiveness}, i.e., the codes of the codebook must cover the variability in the data and no aspect that is relevant for the research question should be left out; and \textit{mutual exclusiveness}, i.e., (i) codes within each domain need to be different and this needs to be clearly specified in the code definitions, and (ii) at most one of the codes of a semantic domain can be applied to a quotation or to overlapping quotations. This means that the codes should have explicit boundaries so that they are not interchangeable or redundant. % We used different colors for each semantic domain and their codes to make the detection of mutual exclusiveness' violations more visible (i.e. no more than one code of the same color can be assigned to a quotation, see Figure~\ref{fig:2}). 

%\begin{figure}[h]
%\centering
%\includegraphics [width=8cm]{figure2.png}
%\caption{Codification in Atlas.ti - Mutual exclusiveness}
%\label{fig:2}
%\end{figure}

The main results of this phase are the list of quotations, which remained unaltered during the subsequent phases, and the first version of the codebook. The codebook is an important finding of this research and the baseline used to answer RQ1 and RQ2. To avoid biases and ensure that the codes mean the same thing to anyone who uses them, it is necessary to build confidence in them. According to Krippendorff, reliability grounds this confidence empirically \cite{Krippendorff:2018} and offers the certainty that research findings can be reproduced: \textit{``The more unreliable the data, the less likely it is that researchers can draw valid conclusions from the data"} \cite{Atlas:2019}. 

{\setlength{\parindent}{0pt} \textbf{Phase 2. Improving the reliability of coding}. We used Inter-Coder Agreement (ICA) analysis techniques to test the reliability of the obtained codebook. This is an iterative process in which Researcher 2 (second author) worked as Coder 1 and Researcher 3 (third author) worked as Coder 2. Coder 1 and Coder 2 were involved to coding the transcriptions of interviews by using a codebook, which was iteratively refined while the ICA did not reach an acceptable level of reliability.  This phase was conducted as follows: Research 1 explained the procedures to Coder1 and Coder 2. This explanation details how to use codes and Atlas.ti projects (e.g. if codes within a semantic domain are not applied in a mutually exclusive manner, the ICA coefficient is inflated and cannot be calculated). After coding, Researcher 4 (fourth author) calculated and interpreted the ICA between Coder 1 and Coder 2. If coders did not reach an acceptable level of reliability, Researcher 1 analyzed the disagreements to to determine why Coder 1 and Coder 2 had not understood a code in the same manner, and delivered a new version of the codebook. Next, Coder 1 and Coder 2 made a new coding on a new subset of interviews. This process was repeated until the ICA reached an acceptable level of reliability.}

This is one of the most important phases for the validity of the study and one of the most complex because, to our knowledge, there is not complete support for ICA statistics in qualitative analysis tools. To address this gap, Researcher 4 used Krippendorff's $\alpha$ coefficients to measure and evaluate the ICA, according to its implementation in Atlas.ti. To accomplish this and to clearly identify the most problematic issues of the codification, a new interpretation of these coefficients was provided in a unified framework.  Section~\ref{method-ica} describes the validity procedure, and Section~\ref{validity} describes its execution.
%(for more information about this framework see Appendices~\ref{appendix-ica} and ~\ref{appendix-alpha}).

{\setlength{\parindent}{0pt} \textbf{Phase 3. Synthesis}. The final product of this step can be a description of higher-order themes, a taxonomy, a model, or a theory. The first action is to determine how many times each domain appears in the data to estimate its relevance (\textit{grounded}). The second action is to support the analysis with evidence through quotations from the interviews. The third action is to calculate a co-occurrence table between the units observed, i.e., between problem domains and between result domains. The fourth action is to create semantic networks, i.e., to analyze the relationships between domains (association, causality, etc.) as well as the relationship strength based on co-occurrence. These relationships determine the \textit{density} of the domains, i.e. the number of domains that are related to each domain. It is possible to repeat these actions for each code within a domain. We did so for more grounded codes.  Finally, it is possible to quantitatively analyze the problems and results by case (organization) or analyze the relationships between problems and results--i.e., interconnecting categories. }

% CREATE A MODEL OF HIGHER-ORDER THEMES.   A theory is a statement of relationships between units observed or approximated in the empirical world.  The first action is to compare fragments from different interviews to which the observers had given a similar code (synonym). This process is known as axial coding, i.e. interconnecting categories [10].  The second action is to look for patterns, which consist in combinations of categories or codes, and to determine if there is any situation that is related with the appearance of a code. This combination is performed using tracing, and hence obtaining clusters/typologies of concepts/codes.

\subsection{Validation procedure}
\label{method-ica}

%We have followed the strategies pointed out by Creswell \cite{Creswell:2002} to improve the validity of exploratory case studies: data triangulation, member checking, rich description, clarify bias, and report discrepant. 
We followed the strategies pointed out by Creswell \cite{Creswell:2002} to improve the validity of exploratory case studies, as follows. 

{\setlength{\parindent}{0pt} \textit{1. Data triangulation} so that the data were gathered from a number of companies that is large enough to build a complete picture of the phenomenon. In this study, 30 companies were included, and 44 stakeholders participated in the interviews. This multiplicity is what provides the basis for \textit{analytical generalization}, where the results are extended to cases that have common characteristics and hence for which the findings are relevant \cite{Wohlin:2012}.}

{\setlength{\parindent}{0pt} \textit{2. Methodological triangulation} so that we used different methods to collect data, i.e., interviews, workshops, observations, and retrieval of a set of metrics. Although this paper focused on the analysis of the interviews, observations and workshop annotations were recorded in a research diary and used to triangulate and complete data omissions.}

{\setlength{\parindent}{0pt} \textit{3. Member checking} so that the participants received the transcribed interview and the preliminary results to ensure the correctness of our findings.}

{\setlength{\parindent}{0pt} \textit{4. Rich description} so that the context of the involved organizations/teams was described as much as confidentiality issues allow. This allowed us to consider \textit{confounding factors} that may have any effect on the outcomes}

{\setlength{\parindent}{0pt} \textit{5. Clarify bias}, i.e., those related to the qualitative research method, such as the bias of the authors, and those related to the participating organizations, such as their location. The first one was mitigated using ICA in the thematic analysis, and the second one was mitigated by the diversity of the interviewees (organizations from different business and industries, in different countries, and from different stakeholders and roles), which increases the generalizability of our results.}

{\setlength{\parindent}{0pt} \textit{6. Report discrepant information} so that all the results are presented and analyzed, regardless of their implications for our initial interests. Prolonged contact with participants, the duration of the interviews, and the subsequent communication allowed us to fully understand their perspectives. }

%Peer debriefing and external auditor, this study are reviewed by other researches, some known directly and others not. 

Of all these strategies, the method used to reduce author bias is especially relevant. As mentioned before, to evaluate the reliability and consistency of the codebook on which the study findings are based, we applied ICA analysis techniques. ICA analysis is a toolbox of widely used statistical methods that provide a formal and standardized way of quantifying the degree of agreement that several judges achieve when evaluating a certain amount of raw material.

Reaching an acceptable threshold of reliability in the codifications is crucial for the validity of the conclusions. Otherwise, logical inference is made based on weak statements without  well-defined and well-bounded codes and domains. This is particularly risky when studying DevOps culture, since the lack of a standardized methodology may lead judges to interpret the same code in different ways. For this reason, ICA is a cornerstone for the validity of the results of this paper.

There exist in the literature a variety of measures for quantifying ICA (c.f.\ \cite{Cohen:1960,Mathet:2015,Fleiss:1971,Scott:1955,Spearman:1904}) that may be applied to different situations. However, for our purposes, we focus on Krippendorff's $\alpha$ coefficient \cite{Hayes:2007,Krippendorff:2004b,Krippendorff:2011,Krippendorff:2016}. This measure is a standard tool for quantifying the agreement in content and thematic analysis due to its well-established mathematical properties and probabilistic interpretations. %There exists a variety of variants of the $\alpha$ coefficient, from which we focus on the following three versions:
In this study, we used the following Krippendorff’s $\alpha$ coefficients: %(see Appendix~\ref{appendix-ica} and Appendix~\ref{appendix-alpha} to see a detail explanation of each one of this coefficients):

\begin{itemize}
\item The coefficient $\alphabin$: This coefficient is computed on a specific semantic domain $S$. It is a measure of the degree of agreement to which coders choose to apply a semantic domain $S$ or not. %A high value of $\alphabin$ is interpreted as an evidence that a domain is clearly stated, its boundaries are well-defined and, thus, the decision of applying it or not is near to be deterministic. However, observe that it does not measure the degree of agreement in the application of the different codes within a domain. Hence, it may occur that the boundaries of a domain are clearly defined but the inner codes are not well chosen. This is not a task of the $\alphabin$ coefficient, but of the cu-a coefficient explained next.

\item The coefficient $\cualpha$: This coefficient is computed on a specific semantic domain $S$. It indicates the degree of agreement to which coders identify codes within $S$.

\item The coefficient $\Cualpha$: In contrast with the previous coefficients, this is a global measure of the goodness of the partition into semantic domains. $\Cualpha$ measures the degree of reliability in the decision of applying the different semantic domains, independently of the chosen code.
\end{itemize}

\begin{figure}[ht]
\centering
\includegraphics [width=12cm]{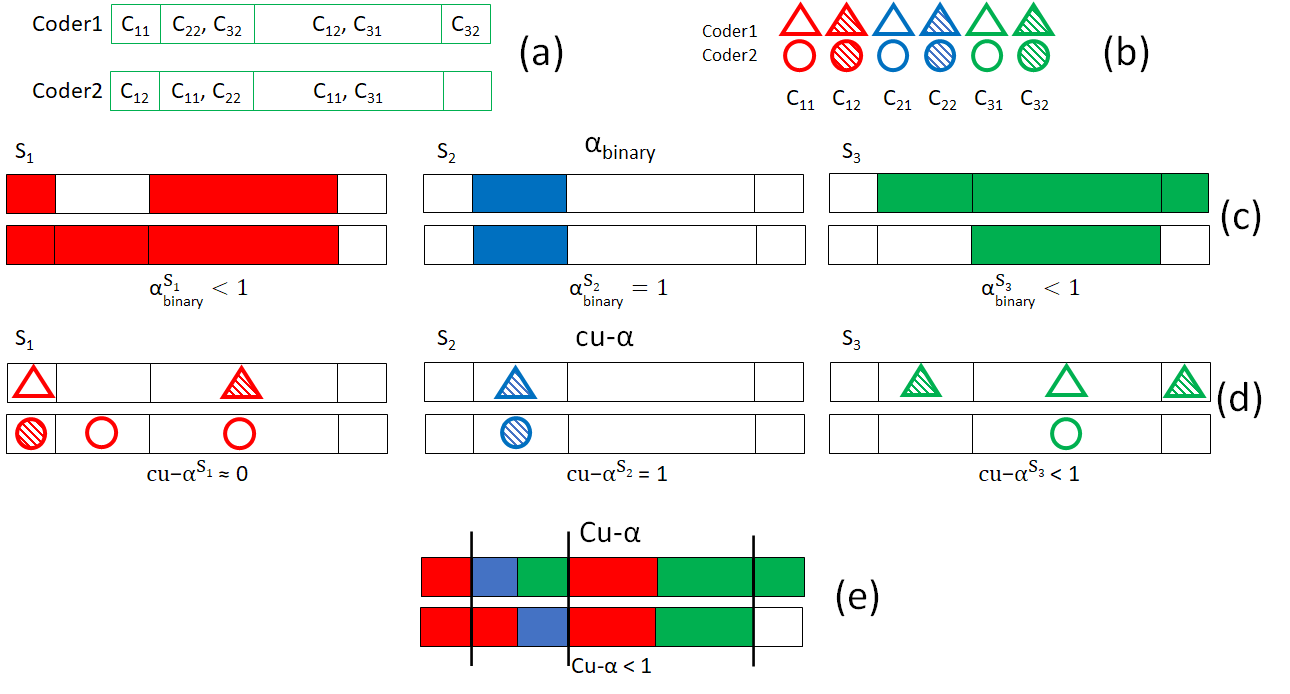}
\caption{Illustrative example for coefficients}
\label{fig:7}
\end{figure}

Figure~\ref{fig:7} shows an illustrative example of the use of these coefficients. Let three semantic domains and their respective codes be:
%$S_1 = \{C_1_1, C_1_2\};$ 		$S_2 = \{C_2_1, C_2_2\}$		$S_3 = \{C_3_1, C_3_2\}$  
$$
	S_1 = \left\{C_{11}, C_{12}\right\}, \quad S_2 = \left\{C_{21}, C_{22}\right\}, \quad S_3 = \left\{C_{31}, C_{32}\right\}.
$$

%This means that, in this running example, we have three semantic domains, $S_1, S_2$ and $S_3$ and each of them subdivides into two codes, $C_{11},C_{12}$ for $S_1$, $C_{21},C_{22}$ for $S_2$, and $C_{31}$ and $C_{32}$ for $S_3$.

Coder 1 and Coder 2 assigned codes to four quotations as shown in Figure~\ref{fig:7}(a), so that the first quotation was assigned ${C_{11}}$ by Coder 1 and ${C_{12}}$ by Coder 2. We created a graphical metaphor so that each coder, each semantic domain, and each code are represented as shown in Figure~\ref{fig:7}(b). Each coder is represented by a shape, such that Coder 1 is represented by triangles and Coder 2 by circles. Each domain is represented by a color: $S_1$ is red, $S_2$ is blue, and $S_3$ is green. Each code within the same semantic domain is represented as a fill, where ${C_{i1}}$ codes are represented by a solid fill and ${C_{i2}}$ codes are represented by dashed fill.

The coefficient $\alphabin$ is calculated per domain (i.e. $S_1$ red, $S_2$ blue, $S_3$ green) and analyzes whether  the coders assigned a domain---independently of the code---to the quotations (see Figure~\ref{fig:7}(c)). Notice that we only focus on the presence or absence of a semantic domain by quotation, so Figure~\ref{fig:7}(c) only considers the color. Now, the $\alphabin$ coefficient measures the agreement that the judges achieved in assigning the same color to the same quotation. The larger the coefficient, the better the agreement. In this way, we obtain total agreement ($\alphabin=1$) for $S_2$ as both coders assigned this domain (blue) to the second quotation, and this domain was absent in the rest of the quotations. On the other hand, $\alphabin<1$ for $S_1$ as Coder 1 assigned this domain (red) to quotations 1 and 3 while Coder 2 assigned it to quotations 1, 2 and 3, leading to a disagreement in quotation 2.

The coefficient $\cualpha$ is also calculated per domain (i.e. $S_1$ red, $S_2$ blue, $S_3$ green), but it measures the agreement attained when applying the codes of that domain. In other words, given a domain $S_i$, this coefficient analyzes whether the coders assigned the same codes of $S_i$ (i.e., the same fills) to the quotations or not. In this way, as shown in Figure~\ref{fig:7}(d), it only focuses on the applied fills to each quotation. In particular, observe that $\cualpha=1$ for $S_2$ since both coders assigned the same code to the second quotation and no code from this domain to the rest of the quotations, i.e., total agreement. Additionally, notice that $\cualpha<1$ for $S_3$ as the coders assigned the same code of  $S_3$ to the third quotation 3 but they did not assign the same codes of  $S_3$ to the rest of the quotations. Finally, observe that the cu-alpha for $S_1$ is very small (near zero) since the coders achieved no agreement on the chosen codes.

Finally, the coefficient $\Cualpha$ analyzes all the domains as a whole, but it does not take into account the codes within each domain. In this way, in Figure~\ref{fig:7}(e), we color each segment with the colors corresponding to the applied semantic domain (regardless of the particular code used). From these chromatic representations, $\Cualpha$ measures the agreement in applying these colors globally among the coders. In particular, notice that $\Cualpha<1$, as both coders assigned the same domain $S_1$ to the first quotations and domains $S_1$ and $S_3$ to the third quotation, but they did not assign the same domains in the second and fourth quotations.

As we mentioned above, the larger the coefficients $\alpha$ are, the better agreement is observed. Typically, the $\alpha$ coefficients lie in the range $0 \leq \alpha \leq 1$. A common rule-of-thumb in the literature \cite{Krippendorff:2018} is that $\alpha \geq 0.667$ is the minimal threshold required for drawing conclusions from the data. For $\alpha \geq 0.80$, we can consider that there exists statistical evidence of reliability in the evaluations.

In Appendix~\ref{appendix-ica} we provide a detailed description of these coefficients with the aim of filling the gap between the multiplicity of descriptions presented in the literature (sometimes too vague) and the current implementation at Atlas.ti (for which only a very brief %and contradictory 
description is provided in the user manual). This Appendix describes a common theoretical framework that allows us to reinterpret the variants of the $\alpha$ coefficient as different incarnations of the same reliability measure. Note that the aim of Appendix~\ref{appendix-ica} is not to formulate new coefficients, but to provide new and unified interpretations in a common framework. In this way, the use of Krippendoff's $\alpha$ does not involve new threats to validity and it is supported by extensive literature. We do not use new coefficients, but give new and clearer interpretations to well-established methods.

Roughly speaking, this measure is computed as the rate of observed disagreement and expected disagreement, following Krippendorff (see Appendix~\ref{appendix-alpha} for a detailed mathematical formulation). The combined methods of these two appendices give rise to a simple and precise algorithm for the computation of these coefficients, as well as the semantics provided above.

\section{Results, Findings, and Discussion}
\label{results}
This section describes the analysis and interpretation of the results for each research question: RQ1 and RQ2. The authors were involved in two iterations of validation and improvement of the codebook related to the problems companies try to solve by implementing DevOps and the results companies try to achieve by implementing DevOps, until an acceptable ICA was reached ($\alpha \geq 0.80$) (see Section~\ref{validity}). A complete version of this codebook in which each code is described in detail is available at GitHub\footnote{\url{https://github.com/jdiazfernandez/DevOpsInPractice/blob/master/codebook.md}, last accessed 01/01/2020}.

%%%%%%%%%%%%%%%%%%%%%%%%%%%%%%%%%%%%%%%%%%%%%%%%%%%%%%%%%%%%%%%%%%%%%%%%%%%%%%%%%%%
%%%%%%%%%%%%%%%%%%%%%%%%%%%%%%%%%%%%%%%%%%%%%%%%%%%%%%%%%%%%%%%%%%%%%%%%%%%%%%%%%%%

\subsection{RQ1: What \textbf{problems} do companies try to solve by implementing DevOps?}
\label{sec:RQ1}

\subsubsection{Codebook}
\label{subsec:codebookproblems}
Table~\ref{tab:CodebookProblems} shows codebook version 2, which lists 10 semantic domains and 35 codes related to the problems that encourage companies to adopt a DevOps culture.  Table \ref{tab:CodebookProblems} also shows how many times each domain and each code appears (i.e., code usage frequency) in the data from the interviews to estimate their relevance (\textit{grounded}). Hence, it is possible to check that P01 \textit{``Too much time for releasing"} is the problem most commonly mentioned by the companies participating in the study (grounded = 27). Next is P10 \textit{``Digital transformation drivers"} (grounded = 15), understood as (i) agile and lean drivers; (ii) movement to DevOps due to client demands, market trends and hypes; and (iii) the need to initiate a transformation due to technological obsolescence or large architectural, infrastructural, and organizational changes. Some other recurring problems are P09 \textit{``Lack of standardization and automation"} (grounded = 14), P02 \textit{``Problems when releasing new versions"} (grounded = 13), P06 \textit{``Organizational and cultural silos"} (grounded = 10), P03 \textit{``Too much time spent on setting up environments"} (grounded = 9) and  P07 \textit{``Lack of collaboration between Dev \& Ops"} (grounded = 9). Each of these codes is described to clarify its meaning through comments in Atlas.ti.

\begin{table}[!ht]
\centering
\scriptsize
\caption{Codebook: problems}
\label{tab:CodebookProblems}
\begin{tabular}{p{0.5cm} p{11cm} p{1.2cm}}
& \\
\hline
\textbf{Id} & \textbf{DOMAIN - Code} & \textbf{Grounded}\\ \hline 
P01 & TOO MUCH TIME FOR RELEASING & 27 \\ 
P01a & Need for being more agile, rapid & 23\\ 
P01b & Need for rapid and continuous feedback loops from ops to dev & 1\\  
P01c & Releases cannot stop production & 2\\ 
P01d & The team does not have not skills for continuous integration and delivery (CI/CD) & 1 \\ \hline 
P02 & PROBLEMS WHEN RELEASING NEW VERSIONS & 13 \\
P02a & Need for higher quality (released) products/services & 7\\ 
P02b & Need for higher quality deployments (into production) & 2\\  
P02c & Releasing problems & 4\\ 
\hline
P03 & TOO MUCH TIME SPENT ON SETTING UP ENVIRONMENTS & 9\\ 
P03a & Operations are not sized for assisting dev requirements (bottleneck) & 2\\  
P03b & Sharing environments by different teams generates conflicts & 1\\  
P03c & Ticketing systems for configuring environments (bottleneck) & 1\\
P03d & Too much time on configuring environments & 5\\ 
\hline 
P04 & SYSTEM DOWNTIME & 2\\  
P04a & System downtime & 2\\ 
\hline 
P05 & BARRIERS TO INNOVATION/EXPERIMENTATION & 2\\  
P05a & The team does not have autonomy (flexibility) to make decisions & 1\\  
P05b & The team has external dependencies to innovate or introduce changes & 1\\ 
\hline
P06 & ORGANIZATIONAL/CULTURAL SILOS & 10\\ 
P06a & Biz \& Dev \& Ops have different goals (business or functional requirements) & 4\\  
P06b & Dev \& Ops have different mindset & 1\\  
P06c & Information/knowledge silos & 3\\ 
P06d & Organizational silos & 2\\
\hline
P07 & PROBLEMS (LACK) OF COLLABORATION BETWEEN DEV \& OPS & 9\\  
P07a & Problems/lack of collaboration/interaction/sync & 2\\ 
P07b & Problems/lack of communication & 5\\  
P07c & Problems/lack of transparency & 2\\ 
\hline
P08 & LACK OF END-TO-END VISION OF VALUE STREAM & 2\\  
P08a & Non-shared (end-to-end) responsibility & 1\\  
P08b & The deployment process is unknown & 1\\ 
\hline 
P09 & LACK OF STANDARDIZATION AND/OR AUTOMATION & 14 \\  
P09a & Complex processes & 1\\ 
P09b & Lack of process automation & 4\\
P09c & Lack of standardized technology stacks, infrastructure, process, methodologies & 3\\
P09d & Lack of version control & 1\\ 
P09e & Need for automating infrastructure creation/configuration & 3 \\ 
P09f & Need for more efficient deployment/production process & 1 \\
P09g & Need for more efficient teams & 1 
\\ \hline
P10 & DIGITAL TRANSFORMATION DRIVERS & 15\\
P10a & Agile \& Lean drivers & 1\\  
P10b & Business/market demands or trends & 7\\  
P10c & Digital transformation or technological obsolescence & 3\\ 
P10d & Large organizational changes & 1\\  
P10e & Large software architectural changes (modernizing legacy applications) & 3\\ 
\hline
\end{tabular}
\end{table}

The relevance can also be measured by the total number of words that a semantic domain accumulated in the transcriptions of the interviews. The analysis provided by Atlas.ti related to the total number of words also suggests that P01, P10, and P02, in this strict order, were the most relevant, while P04, P05, and P08 were the least relevant.

Next, we analyze the most relevant codes, i.e., the most frequently used codes (grounded, according to Atlas.ti nomenclature) that show the problems that instilled a DevOps culture, and we provide evidence for each code through excerpts of the transcriptions: 

%\begin{figure}[!h]
%\centering
%\includegraphics [width=10cm, frame]{figure3.png}
%\caption{Codebook version 2}
%\label{fig:figure3}
%\end{figure}

P01a \textit{``Need for being more agile, rapid"} is the most relevant code, which is mentioned up to 23 times (grounded = 23). Hence, the organizations claimed that the most important problem was the amount of time required for releasing new features into production. They needed to reduce time-to-market and be more agile/rapid to adapt to market needs and demands, either new features or updates. Business demands more velocity than teams can offer. Hence, the organizations emphasized that they took too much time to deploy, deliver, and release new features and hotfixes and talked about the need to accelerate the \textit{``value"} delivery to customers. P01a is evidenced by the following excerpts: \textit{``The time between when development teams finished a new software and when it was deployed into production was very long, sometime months''} [ID01], \textit{``The main concern is the time it takes to put a release into production. Without emphasizing the origin of the problem, the client told us that it takes a long time to put into production the requested changes, and that they want to adopt DevOps to reduce this time''} [ID06], \textit{``The speed demanded by business was not according to the speed provided by operations. It took us a long time to develop features and when they went into production, there was something else new. This way of working does not allow us to iterate and adapt to market needs quickly''} [ID11], \textit{``The client had very traditional developments based on waterfall methodologies and it was very difficult for them to respond to the business needs. Every time they wanted to make a change and put that change into production, it took approximately 3 or 4 months''} [ID19], \textit{``We were late, frequently, we could not deploy into production because software applications were not working or hotfixes were being solved. When we solved an operating error, four more errors appeared later''} [ID16], and finally, \textit{``The most important driver is that we were unable to generate a hotfix quickly and over-the-air"} [ID22].

P10b, \textit{``Business/market demands or trends''} (grounded = 7), is related to the digital transformation drivers mentioned above. This indicates that many times the reasons for adopting DevOps do not come from within the organization, but from outside, such as customer and market requests. P10b is evidenced by the following excerpts: \textit{``We started adopting DevOps because of customer demands''} [ID08], \textit{``The clients we work with were starting a DevOps transformation because the trend that is being sold from the market is that with the introduction of DevOps you are going to get a higher quality software in less time. The trend has a lot of influence''} [ID16], \textit{``We haven't adopted DevOps as a resolution to a problem, but as a need to adapt to a market with enormous variability that requires much more agility than we were capable of providing"} [ID04], and finally, \textit{``The market is asking for DevOps. We came from applying traditional development models and were not aware of the market situation and technological evolution. We had to make a change within the organization and adapt to a new model''} [ID14].

P02a \textit{``Need for higher quality (released) products/services''} (grounded = 7). This means that the organizations participating in the study need velocity, but they also needed high-quality products and services, i.e., they needed to release software with the least number of bugs as possible and improve the user experience and other quality metrics, such as performance. P02a is evidenced by the following excerpts: \textit{``In the end, the problem was faster value delivery but also higher quality delivery and quicker problem solving''} [ID30], \textit{``Production deployments needed a lot of speed and quality''} [ID09], and finally, \textit{``The performance of the service was very limited, from a certain number of users connected, navigation on the platform became slow and unstable. We needed to improve the user experience (browsing speed, minimize performance drops,...) and optimize hardware resources''} [ID07].

P03d \textit{``Too much time on configuring environments''} (grounded = 5). It is striking that large and important companies have such problems when configuring their environments. P03d is evidenced by the following excerpts: \textit{``Setting up new environments or updating existing environments was very time consuming. In addition, different teams shared environments so it was very difficult to agree on the changes that affected the environments in which the different developments were deployed"} [ID01]. \textit{``You asked for some machines and the IT department took two months; you never knew when your request would be resolved (maybe there were thousands of requests in such a large company). After two months the machines were delivered, but without IP, and you entered a list of requests again to request the IP. We weren't well organized and each team was very slow"} [ID02].

P07b \textit{``Problems/lack of communication''} and lack of understanding between development and operations (grounded = 5). Software products and services are built or provided through the interaction and collaboration of very diverse professionals. If these professionals do not communicate adequately the timing and quality of the product/service will be compromised. P07b is evidenced by the following excerpts: \textit{``There was a lack of communication between the operations and development engineers, which meant that they did not understand each other"} [ID01], \textit{``One of the problems we encountered was lack of communication between the developers and deployment teams. The deployment team was completely unaware of what was being implemented"} [ID12], \textit{``There was also a lot of distance, in terms of understanding, between what people developed and wanted to deploy and how they deployed it. There was not much communication with operations and they were not even prepared to meet the deployment requirements"} [ID11], and finally, \textit{``In the end, the lack of collaboration between development and operations caused many delivery failures''} [ID10].

P06a \textit{``Biz \& Dev \& Ops have different goals (business or functional requirements)''} (grounded = 4), which discloses the problem caused by the goals misalignment of the parties involved in the delivery of a software product/service, i.e. business, development and operations. Hence, operations personnel do not inform developers  about operational requirements or changes, while operations does not receive sufficient information about how systems work and does not have sufficient understanding of the business. Some organizations participating in the study claimed that \textit{``There were barriers between development and operations teams as they did not share common objectives. Development was focused on implementing new functionality and operations in maintaining the system stable"} [ID26], \textit{``On the one hand, developers want to deploy more often, because they want to bring their software into production. On the other hand, system engineers are looking for application stability, and changes destabilize"} [ID01], and \textit{``Development worked very close to business, but very far from operations. Therefore, the software functionally fulfilled most of the requirements of the business, but the software did not meet the operational requirements. The system sometimes went down because millions of people were simultaneously accessing to it and no one had thought about it"} [ID25]. 

P09b \textit{``Lack of processes automation''} (grounded = 4) and P09e \textit{``Need for automating infrastructure creation and configuration''} (grounded = 3) are related to the need for automated processes for building, testing, integrating, deploying, and releasing software and the need for automating infrastructure (infrastructure as code, immutable infrastructure, etc.). Some organizations revealed that manual deployments generate problems and failures and are associated with a high cost. These codes are evidenced by the following excerpts:  \textit{``The manual deployments generated failures from time to time; in fact, there were some manual errors that generated important problems''} [ID22], \textit{``Millions of euros were spent on deployments, which were manual''} [ID25], \textit{``The lack of automation in the processes slowed down the releasing time a lot."} [ID10], \textit{``We are going to be a company 24/7 and we need releases to not stop production, so we need a quality deployment pipeline with a lot of automation and control"} [ID09], and \textit{``We needed to have automation of the infrastructure of the environments, dynamic growth of it, dynamic provisioning, and that’s when we saw that we needed DevOps''} [ID05].

Finally, it is also noteworthy that the larger the companies are, the larger the organizational problems are. Hence, some organizations participating in the study revealed that \textit{``This organization is very large and complex. Development and infrastructure were silos that made successful release very difficult"} [ID02]. \textit{``We realized that the very size of the company, the diversity of departments (development, operations, security, service, QA, architecture, etc.), and the complexity of the processes, as well as the interaction between them, went against reducing time-to-market, and made us less competitive"} [ID17]. These excerpts are related to P06d \textit{``Organizational silos''}, although there also exist P06c \textit{``Information/knowledge silos''} that do not explicitly respond to organizational silos, such as departments. Additionally, some organizations revealed that operations are not sized for assisting development requirements (P03a): \textit{``The operations personnel were overwhelmed by all the resources they had to provide to development teams so that  they could not improve their tasks. This was the primary bottleneck in the software development process''} [ID13].   % Indeed, improving their ability to compete is precisely what has driven many companies to adopt DevOps  \textit{``With the new needs of bringing out new developments in mobile applications against the competition, a new DevOps team oriented to new technologies and to accelerate the production of those new developments was created''} [ID19].

%%%%%%%%%%%%%%%%%%%%%%%%%%%%%%%%%%%%%%%%%%%%%%%%%%%%%%%%%%%%%%%%%%%%%

\subsubsection{Semantic Networks}
\begin{figure}[!h]
\centering
\includegraphics [height=18cm]{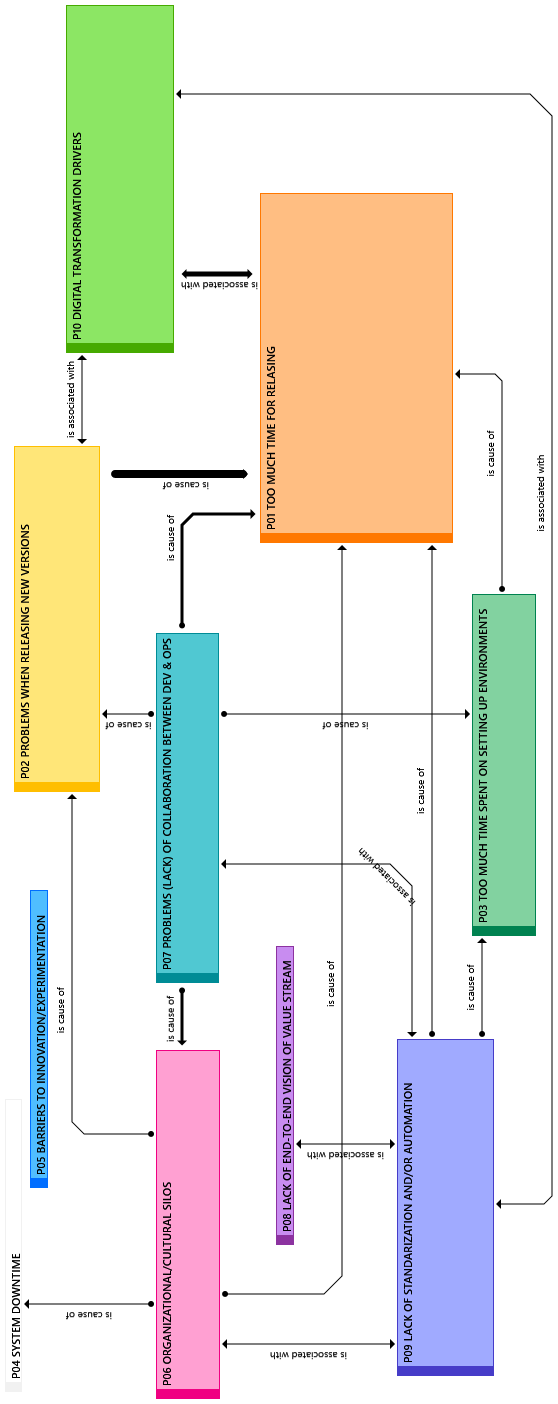}
\caption{Semantic network for problems' domains}
\label{fig:NetworkProblems}
\end{figure}

The abovementioned domains and codes are related to each other through semantic networks as described next. Figure~\ref{fig:NetworkProblems} shows a semantic network about the problems that lead organizations to move to DevOps and how these problems are related to each other. An analysis of co-occurrences among the semantic domains is used to establish relations between domains and to understand how strong these relations are (see Table~\ref{tab:CoocurrencesProblemsDomain}). Each cell of this table shows the co-occurrence of two semantic domains in a quotation. In the network, these values determine the width of the line connecting the domains, i.e., the higher the co-occurrence is, the thicker the line. Furthermore, the network establishes the type of  relations between two domains. In this network, two types of relationships are identified: \textit{is associated with}, which is generic, and \textit{is cause of}, which indicates causality (but in an inductive, not statistical, sense). Finally, the size of the boxes (specifically the value of the height) containing the semantic domains indicates their grounded value, i.e., the more strongly grounded the domain is, the taller the box. These considerations must be taken to properly understand the semantic network. Next, we analyze the network from the most to the least grounded domains, as follows:

\begin{table}[!h]
\centering
\small
\caption{Co-occurrences of problems at the domain level \tablefootnote{Note that, for simplicity, we represent the occurrence relation only in the upper-diagonal part}}
\begin{tabular}{p{0.6cm} p{0.6cm} p{0.6cm} p{0.6cm} p{0.6cm} p{0.6cm} p{0.6cm} p{0.6cm} p{0.6cm} p{0.6cm} p{0.6cm} }
& \\
\hline \hline
\textbf{ } & \textbf{P01} & \textbf{P02} & \textbf{P03} & \textbf{P04}& \textbf{P05}& \textbf{P06}& \textbf{P07}& \textbf{P08}& \textbf{P09}& \textbf{P10}\\ \hline 
\textbf{P01} & - & 9 & 1 & 0 & 0 & 2 & 3 & 0 & 2 & 5\\ 
\textbf{P02} & - & - & 0 & 0 & 0 & 1 & 1 & 0 & 0 & 2\\ 
\textbf{P03} & - & - & - & 0 & 0 & 0 & 2 & 0 & 1 & 0\\  
\textbf{P04} & - & - & - & - & 0 & 1 & 0 & 0 & 0 & 0\\  
\textbf{P05} & - & - & - & - & - & 0 & 0 & 0 & 0 & 0\\  
\textbf{P06} & - & - & - & - & - & - & 3 & 0 & 1 & 0\\  
\textbf{P07} & - & - & - & - & - & - & - & 0 & 1 & 0\\  
\textbf{P08} & - & - & - & - & - & - & - & - & 1 & 0\\ 
\textbf{P09} & - & - & - & - & - & - & - & - & - & 2\\ \hline 
\hline
\end{tabular}
\label{tab:CoocurrencesProblemsDomain}
\end{table}

%\begin{landscape}
%\end{landscape}

Problem P01, \textit{``Too much time for releasing"} (grounded = 27), has a strong relationship with P02 \textit{``Problems when releasing new versions"} (co-occurrence = 9). From these data, we could induce that releasing problems may \textit{cause} delivering software to take too much time. Finally, P01 is related to P07 \textit{``Lack of collaboration between dev \& ops"}, P09 \textit{``Lack of standardization and/or automation"}, P06 \textit{``Organizational/cultural silos''}, and P03 \textit{``Too much time spent on setting up environments"}, likely \textit{causing} the excessive time that companies invest in releasing software.  

Problem P10, \textit{``Digital transformation drivers''} (grounded = 15), has a strong \textit{association with} P1 (co-occurrence = 5). P10 is \textit{associated with} P02 and P09. Thus, it seems that the too much time for releasing, the problems when releasing, and the lack of standardized and automated processes, technological stacks, infrastructure, etc., drive organizations to implement a digital transformation. 

Problem P09, \textit{``Lack of standardization and automation''} (grounded = 14), is \textit{associated with} P06, i.e., the existence of silos, either organizational or cultural silos, and P07, i.e., the lack of collaboration between development and operations teams. It seems that the lack of collaboration and the existence of silos make it difficult or even impossible to standardize of methods, tools, and technologies and to automate of processes. 

Problem P02, \textit{``Problems when releasing new versions''} (grounded = 13), is related to P01 and P10, and is also related to P06 and P07. Thus, it seems that organizational and cultural silos, as well as the lack of collaboration between dev and ops, may \textit{cause} problems when releasing new versions.  

Problem P06, \textit{``Organization and cultural silos''} (grounded = 10), has a strong relationship with P07. It seems that the lack of collaboration between dev and ops \textit{causes} these organization and cultural silos in the companies participating in the study. In turn, organization and cultural silos may \textit{cause} P04, i.e., systems' downtimes. 

Finally, P03, \textit{``Too much time spent on setting up environments''} (grounded = 9), is related to P07 and P09. Thus, it seems that P03 is caused by P07 and P09, i.e., by the lack of collaboration and the lack of standardization and automation, respectively. 

\begin{figure}[!h]
\centering
\includegraphics [height=15cm]{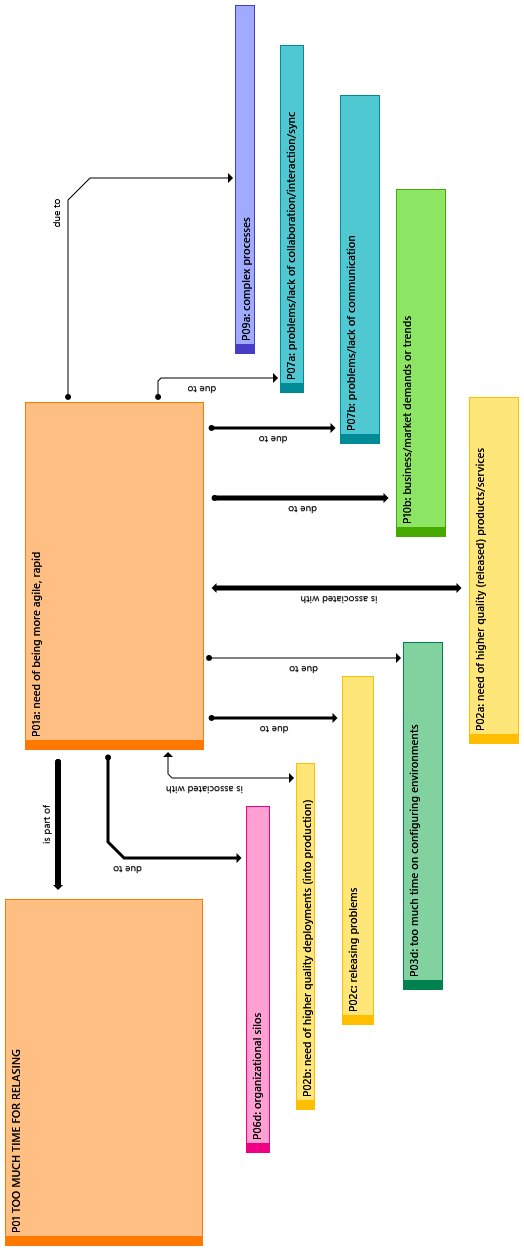}
\caption{Semantic network for problems' codes}
\label{fig:NetworkProblemP01a}
\end{figure}

In addition to the semantic network at the domain level of Figure~\ref{fig:NetworkProblems}, Figure~\ref{fig:NetworkProblemP01a} shows a semantic network at the level of codes, specifically around the most grounded and, hence, the most outstanding code. This network is based on the co-occurrences between this code and others (see Figure~\ref{fig:CoocurrenceTableP01a}). As stated above, the most grounded code is P01a, \textit{``Need for being more agile, rapid"} (grounded = 23), and this is the base code of this network. The network is created following the guidelines mentioned above regarding the lines and boxes. The network relates the base code with other elements through three kinds of relationships: \textit{is part of}, which links the code to its domain, \textit{is associated with}, which is generic, and \textit{due to}, which indicates causality (but in an inductive, not-statistical sense). Hence, Figure~\ref{fig:NetworkProblemP01a} shows that P01a has a strong relationship with P02a \textit{``Need for higher quality products"} and P10b \textit{``Business and market demands or trends"} (co-occurrence = 5). This means that the organizations that mentioned P01a in a quotation, also mentioned these codes, so agility, quality, and business/market demands and trends are intrinsically linked. P01 has a medium relationship with  P06d \textit{``Organizational silos"}, P02c \textit{``Releasing problems"}, and P07b \textit{``Lack of communication"} (co-occurrence = 2). Thus, it seems that the need to be more agile may be, to some extent, due to these problems. Finally, P01a is weakly related to P03d \textit{``Too much time on configuring environments"}, P09a \textit{``Complex processes"}, and P02b \textit{``Need for higher quality deployments"} (co-occurrence = 1).

%\begin{landscape}
%\end{landscape}

\begin{table}[!h]
\centering
\small
\caption{P01a co-occurrence table}
\begin{tabular}{p{9cm} p{4.3cm}}
& \\
\hline \hline
 & \textbf{P01a} need of being more agile, rapid \\ \hline 
\textbf{P02a} need for higher quality (released) products/services &  5\\ 
\textbf{P02b} need for higher quality deployments (into production) &  1\\ 
\textbf{P02c} releasing problems &  2\\  
\textbf{P03d} too much time on configuring environments &  1\\  
\textbf{P06d} organizational silos  &  2\\  
\textbf{P07a} problems/lack of collaboration/interaction/sync&  1\\  
\textbf{P07b} problems/lack of communication &  2\\  
\textbf{P09a} complex processes &  1\\ 
\textbf{P10b} business/market demands or trends &  5\\ \hline 
\hline
\end{tabular}
\label{fig:CoocurrenceTableP01a}
\end{table}

%%%%%%%%%%%%%%%%%%%%%%%%%%%%%%%%%%%%%%%%%%%%%%%%%%%%%%%%%%%%%%%%%%%%%%%%%%%%%%%%%%%
%%%%%%%%%%%%%%%%%%%%%%%%%%%%%%%%%%%%%%%%%%%%%%%%%%%%%%%%%%%%%%%%%%%%%%%%%%%%%%%%%%%

\subsection{RQ2: What \textbf{results} do companies try to achieve by implementing DevOps?}
\label{sec:RQ2}

\subsubsection{Codebook}
\label{subsec:codebookresults}

Table~\ref{tab:CodebookResults} shows the codebook that lists 6 semantic domains and 18 codes related to the results and benefits that companies expect to achieve as a result of adopting DevOps. Table \ref{tab:CodebookResults} also shows how many times each domain and each code appears (i.e., code usage frequency) in the data from the interviews to estimate their relevance (\textit{grounded}). Hence, it is possible to see that R01 \textit{``Faster time-to-market"} is the result most commonly mentioned by the companies participating in the study (grounded = 27); thus, it seems to be the most important result that organizations expect when initiating the adoption of the DevOps culture. This result is followed by other relevant results, such as R03 \textit{``Improve process productivity''} (grounded = 20), R02 \textit{``Better software quality''} (grounded = 19), and R04 \textit{``Improve team effectiveness \& satisfaction''} (grounded = 13). Other expected results expressed to a lesser extent by the companies are R05 \textit{``Customer focus''} (grounded = 7) and R06 \textit{``Align the objectives of Dev \& Ops with Business''} (grounded = 2).

%As presented in the previous section, the authors were involved in two iterations of improvement of the codebook related to the results that companies try to achieve by implementing DevOps, until an acceptable ICA was reached ($\alpha \geq 0.80$).

\begin{table}[b]
\centering
\small
\caption{Codebook: Results}
\begin{tabular}{p{0.8cm} p{10.5cm} p{1.2cm}}
& \\
\hline
\textbf{Id} & \textbf{DOMAIN - Code} & \textbf{Grounded}\\ \hline 
R01 & FASTER TIME-TO-MARKET & 27 \\ 
R01a & Agile response to customers/market & 4 \\ 
R01b & Fast and continuous integration & 1 \\ 
R01c & Faster response of hotfixes & 3 \\
R01d & Faster time-to-market & 19 \\ \hline
R02 & BETTER SOFTWARE QUALITY  & 19 \\ 
R02a & Better software quality & 15 \\ 
R02b & Reliability, resilience (recoverability), availability & 4 \\ \hline
R03 & IMPROVE PROCESS PRODUCTIVITY & 20  \\ 
R03a & Process automation: efficiency, optimization, productivity & 15 \\
R03b & Project management more effective & 2 \\
R03c & Reduce IT cost & 3 \\ \hline
R04 & IMPROVE TEAM EFFECTIVENESS \& SATISFACTION & 13 \\ 
R04a & Build trust within the team and between silos & 2 \\ 
R04b & Improve team autonomy/flexibility & 2 \\ 
R04c & Improve team collaboration $\&$ communication & 8 \\ 
R04d & More motivated teams & 1 \\ \hline
R05 & CUSTOMER FOCUS & 7 \\ 
R05a & Experimentation/innovation & 2 \\ 
R05b & Fast and continuous feedback & 2 \\ 
R05c & Greater value to customers (excellence) & 3 \\ \hline
R06 & ALIGN THE OBJECTIVES OF DEV \& OPS WITH BIZ & 2 \\ 
R06a & Align objectives with business & 1 \\ 
R06b & End-to-end vision of value stream & 1  \\ \hline
\end{tabular}
\label{tab:CodebookResults}
\end{table}

The relevance can also be measured by the total number of words a semantic domain accumulated in the transcriptions of the interviews. The analysis provided by Atlas.ti related to the total of words also supports that R03, R01, R04, and R02, in this strict order, were the most relevant, while R05 and R06 were the least relevant (see the study repositories).

Next, we analyze the most relevant codes, i.e., the most frequently used codes (grounded, according to Atlas.ti nomenclature) that show the results that organizations expect to achieve when adopting a DevOps culture, and we provide some evidence for each code through excerpts of the transcriptions: 

R01d \textit{``Faster time-to-market''} is the most relevant code, which is mentioned up to 19 times (grounded = 19). Hence, the organizations reported that the most expected benefit when they initiated their DevOps transformation was the acceleration of the time-to-market by reducing development, testing, quality assurance, deployment and delivery times. R01d is evidenced by the following  excerpts: \textit{``It was expected that DevOps enables the development of new functionalities and their deployment into production for final users more quickly''} [ID21], \textit{``We expected that DevOps would allow us to give a faster response to the client. Currently agility is an absolute necessity''} [ID04], and \textit{``By incorporating DevOps we will gain agility and achieve a better adaptation to the current changing market''} [ID14].

R02a \textit{``Better software quality''} (grounded = 15). The organizations also reported important expected results and benefits related to the increase in software quality. These results are supported by the following excerpts:  \textit{``We expected an improvement in the quality of the deliverables''} [ID14], \textit{``It was expected that DevOps technologies would support us to deliver software faster, but also to build software with better quality. In the end, you have thousands and thousands of automated tests that are going to be executed for a new code that you want to deploy''} [ID03], and \textit{``We are looking for productivity and efficiency as well as quality in software. These results facilitate what top management wants: fast time-to-market and competitiveness''} [ID05]. 

R03a \textit{``Process automation: efficiency, optimization, productivity''} (grounded = 15). R03a is evidenced by the following  excerpts: \textit{``We will be able to deploy at any time"} [ID22], \textit{``We aim for an automated model in the provisioning of the platform''} [ID27] and \textit{``We expected the benefits of automating tasks, previously done manually, through continuous integration and continuous deployment. This avoids human mistakes and shortens time to production''} [ID24]. In this regard, some striking issues that organizations expect to eliminate by adopting DevOps are described through the following excerpt: \textit{``There were several manual changes made by particular developers and this caused several problems. Sometimes these developers were on vacation and no one knows what they touched. Sometimes, a change was made by hand in production and the next time something was uploaded and overwritten, the change was lost and no one remembered how to do it''} [ID28].

R04c \textit{``Improve team collaboration and communication''}  (grounded = 8). R04c is evidenced by the following  excerpts: \textit{``The aim of incorporating DevOps was to improve communication and dialogue between all parties involved in the projects''} [ID12], \textit{``We were looking to break down all the barriers that initially existed between the development and operations teams so that everyone shares common objectives''}[ID26], \textit{``We were looking for a satisfactory process for development and operations to generate more confidence. Operations did not trust that the developers were able to deploy in production; meanwhile, developers were always criticizing that they were not allowed to deploy''} [ID01], and \textit{``The main result we were expecting was the reduction of the distance between the development and operations teams. We do not want to spend time deploying code, integrating code, solving problems in different branches, etc. We want to spend time on design, development, learning about new technologies, innovation, etc. This makes the company much more competitive''} [ID24].

Finally, the organizations also highlighted that they expected that software products would be more reliable and resilient (R02b), the management of the projects would be more effective (R03b), and the IT cost would be reduced (R03c). This can be observed in excerpts such as these: \textit{``We expected an improvement in response times to incidents, performance and communication of the systems that support the platform, as well as more effective project management''} [ID07], \textit{``As a result of changing traditional project management to product-oriented management, we expected to improve the management and shorten development, deployment and QA times''} [ID12], \textit{``We wanted to bring the operations teams closer to the development and business teams. That is, if business asked for a new website, our idea was to give an end-to-end vision of this product. We did not want development and operations working separately; we wanted everything to be a single stream leading up to a new website''} [ID02], and \textit{``In the end, the main role of management is to seek fast time-to-market and decrease IT costs''} [ID05].

Once the most grounded codes were analyzed---and in turn, the most relevant domains R01-R04---, we focused on the codes framed in R05~\textit{``Customer focus''} (grounded = 7). The companies also reported some expected results related to customer-centered approaches. The organizations claimed that by introducing DevOps, they expected to manage customers' responses faster, deliver greater value, obtain feedback more quickly and continuously, and innovate more. These expected outcomes can be found in the following excerpts: \textit{``Above all, we expect to provide greater value to customers and get higher satisfaction''} [ID08], \textit{``Improve the user experience, faster feedback from the end user so that the business gets that feedback''} [ID11], \textit{``By making cycles faster, we aim to get better customer feedback and expect that customers get what they truly want and that the user experience is better too''} [ID10], and finally, \textit{`DevOps makes innovation less scary. We expect to spend more time on new things because pipelines break less so that you do not have the experience of production breaking down for several days because you have tried something new''} [ID03].

%%%%%%%%%%%%%%%%%%%%%%%%%%%%%%%%%%%%%%%%%%%%%%%%%%%%%%%%%%%%%%%%%%%%%
\subsubsection{Semantic Networks}

\begin{figure}[!h]
\centering
\includegraphics [height=14.5cm]{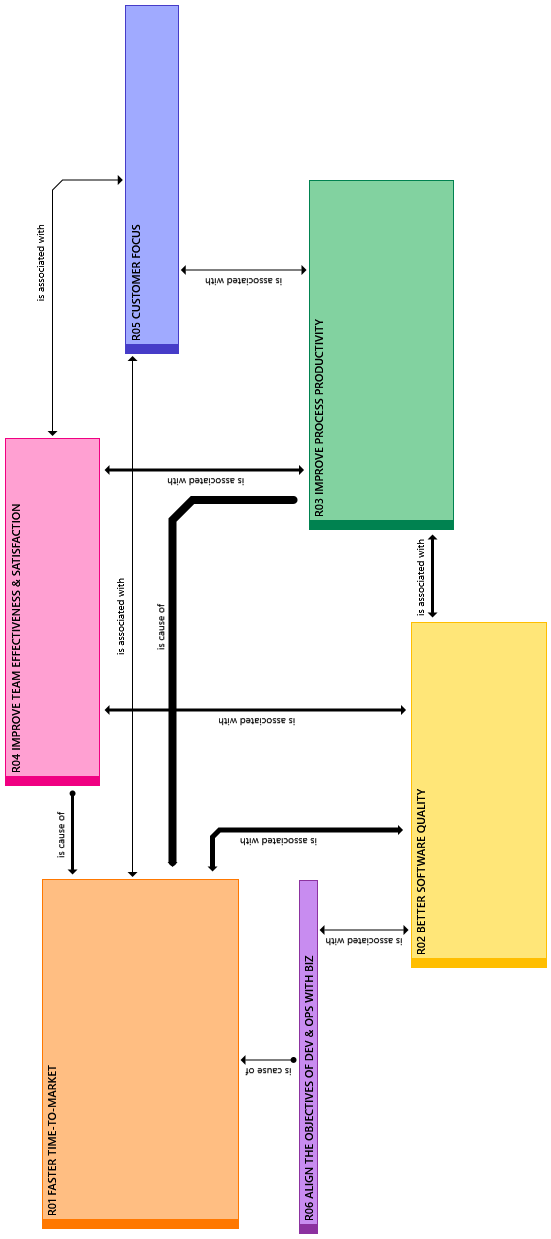}
\caption{Result's semantic network}
\label{fig:NetworkResults}
\end{figure}

The abovementioned domains and codes are related to each other through semantic networks, as described next. Figure~\ref{fig:NetworkResults} depicts a network of the results that organizations expected to achieve when they initiated the DevOps adoption and how these results are related to each other. Table~\ref{tab:CoocurrencesResultsDomain} presents the co-occurrences analysis at the domain level that enables the creation of the network. These networks are created following the guidelines previously mentioned: the height of the boxes is determined by the grounded value of the elements, the relations between them are based on the analysis of co-occurrences, the co-occurrence values determine the line width that links the elements, and several types of relations characterize these relationships. Next, we analyze the network from the most to the least grounded domains, as follows:

\begin{table}[t]
\centering
\small
\caption{Co-occurrences of the results at the domain level \tablefootnote{Note that, for simplicity, we represent the co-occurrence relation only in the upper-diagonal part}}
\begin{tabular}{p{0.6cm} p{0.6cm} p{0.6cm} p{0.6cm} p{0.6cm} p{0.6cm} p{0.6cm} p{0.6cm} p{0.6cm} p{0.6cm} p{0.6cm}}
& \\
\hline 
\hline 
\textbf{ } & \textbf{R01} & \textbf{R02} & \textbf{R03} & \textbf{R04}& \textbf{R05}& \textbf{R06}\\ \hline 
\textbf{R01} & - & 7 & 10 & 4 & 1 & 1 \\ 
\textbf{R02} & - & - & 4 & 3 & 0 & 1 \\ 
\textbf{R03} & - & - & - & 3 & 1 & 0 \\ 
\textbf{R04} & - & - & - & - & 1 & 0\\ 
\textbf{R05} & - & - & - & - & - & 0 \\ 
\hline
\hline 
\end{tabular}
\label{tab:CoocurrencesResultsDomain}
\end{table}

%\begin{landscape}
%\end{landscape}

Result R01, \textit{``Faster time-to-market''} (grounded = 27), also has the largest number of relationships, and these relations are the strongest. Therefore, it could be inferred that this is the most relevant result that companies intend to achieve when adopting DevOps. R01 has a strong relationship with  R03  \textit{``Improve process productivity''} (co-occurrence = 10). Thus, it seems that the optimization and automation of processes may \textit{cause} faster time-to-market. R01 is also strongly \textit{associated with} R02  \textit{``Better software quality''} (co-occurrence = 7), which shows that organizations expect to achieve both speed and quality. R01 is also related to R04 \textit{``Improve team effectiveness \& satisfaction''} (co-occurrence = 4); thus, it seems that, to accelerate time-to-market, not only the effectiveness of process but also people's satisfaction and team effectiveness are important. Finally, R01 is weakly related to R05 \textit{``Customer focus''} (co-occurrence = 1) and R06 \textit{``Align the objectives of Dev \& Ops with Biz''} (co-occurrence = 1).

Result R03,  \textit{``Improve process productivity''} (grounded = 20), is associated with R04 (co-occurrence = 3) and R05 (co-occurrence = 1), showing that processes, people \& teams, and customers are intrinsically related. %R03 is also strongly related to R02 (co-occurrence = 4), so it seems that the improvement of processes productivity may lead to better software quality.

Result R02,  \textit{``Better software quality''} (grounded = 19), is strongly related to the semantic domains R01 (co-occurrence = 7), R03 (co-occurrence = 4) and R04 (co-occurrence = 3), and weakly related to R06 (co-occurrence = 1). In a nutshell, it seems that the adoption of the DevOps culture in software companies can favor higher levels of process productivity and team effectiveness and satisfaction, and in turn, it may lead to improved software quality.

Finally, R04, \textit{``Improve team effectiveness \& satisfaction''}, is related to the abovementioned R01, R02, R03 and associated with R05 in such a way that the improvement in team effectiveness and people's satisfaction is related to the improvement in customer satisfaction and experience by providing excellent service to customers. 

\begin{figure}[!h]
\centering
\includegraphics [height=18cm]{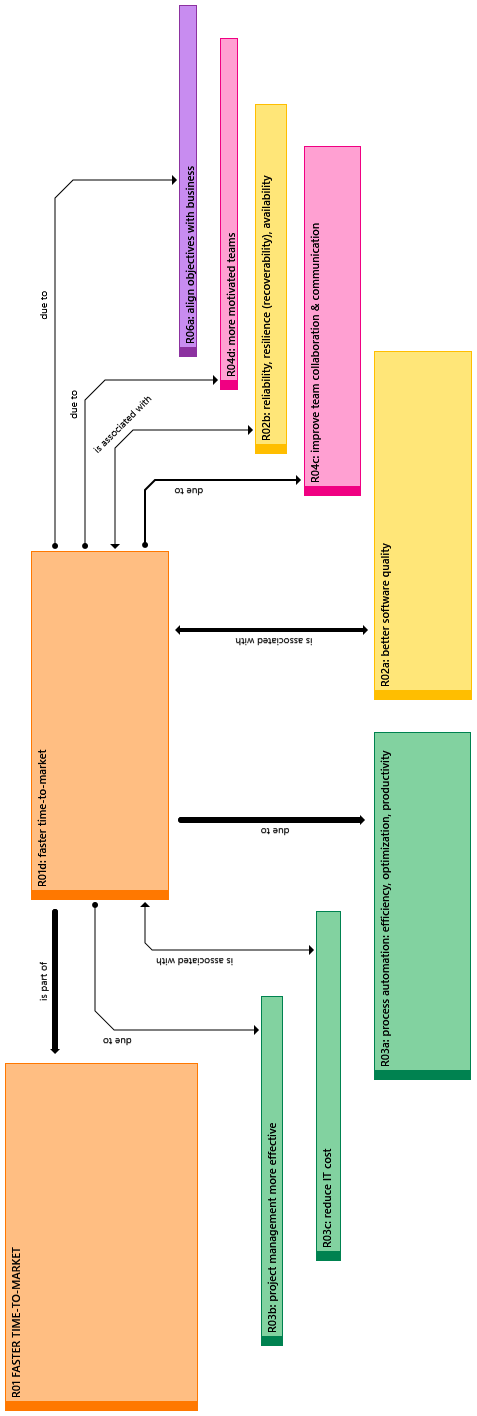}
\caption{R01d's semantic network results}
\label{fig:NetworkResultR01d}
\end{figure}

In addition to the semantic network at the domain level of Figure~\ref{fig:NetworkResults}, Figure~\ref{fig:NetworkResultR01d} shows a semantic network at the level of codes, specifically around the most grounded and thus outstanding code. This network is based on the co-occurrences between this code and others (see Figure~\ref{fig:CoocurrenceTableR01d}). As stated before, the most grounded code is R01d \textit{``Faster time-to-market''} (grounded = 19), which is the base code of this network. The network is created following the guidelines mentioned above regarding the lines and boxes. The network relates the base code with other elements through three kinds of relationships: \textit{is part of}, which links the code to its domain, \textit{is associated with}, which is generic, and \textit{due to}, which indicates causality (but in an inductive, not-statistical, sense). Hence, Figure~\ref{fig:NetworkResultR01d} shows that R01d has a strong co-occurrence with R03a \textit{``Process automation: efficiency, optimization, productivity''} (co-occurrence = 7), which is also very relevant (grounded = 15). This means that the organizations that mentioned R01d in a quotation also mentioned R03a. Thus, it seems that faster time-to-market is largely due to process automation. R01d is also strongly associated with R02a \textit{``Better software quality''} (co-occurrence = 4), confirming that organizations expect to obtain both fast speed and high software quality. R01d is also related to R04c \textit{``Improve team collaboration $\&$ communication''} (co-occurrence = 2), and henceforth, it could be inferred that the increase in velocity may also be due to the improvement in team collaboration. Other codes weakly related to R01d are R04d \textit{``More motivated teams''}, R06a \textit{``Align objectives with business''} and R03b \textit{``Project management more effective''} (co-occurrence = 1). Thus, it seems that the faster time-to-market that companies aim to achieve could also be due to these other expected results. 

%\begin{landscape}
%\end{landscape}

\begin{table}[!h]
\centering
\small
\caption{R01d co-occurrence table}
\begin{tabular}{p{9cm} p{3.5cm}}
& \\
\hline \hline
 & \textbf{R01d} faster time-to-market \\ \hline 
\textbf{R02a} better software quality &  5\\ 
\textbf{R02b} reliability, resilience (recoverability), availability &  1\\ 
\textbf{R03a} process automation: efficiency, optimization, productivity &  2\\  
\textbf{R03b} more effective project management &  1\\  
\textbf{R03c} reduce IT cost &  2\\  
\textbf{R04c} improve team collaboration \& communication &  1\\  
\textbf{R04d} more motivated teams &  2\\  
\textbf{R06a} align objectives with business &  1\\ 
\hline
\end{tabular}
\label{fig:CoocurrenceTableR01d}
\end{table}

%%%%%%%%%%%%%%%%%%%%%%%%%%%%%%%%%%%%%%%%%%%%%%%%%%%%%%%%%%%%%%%%%%%
%%%%%%%%%%%%%%%%%%%%%%%%%%%%%%%%%%%%%%%%%%%%%%%%%%%%%%%%%%%%%%%%%%%
\subsection{Findings and Discussion}
This section interprets the semantic networks illustrated in Figures~\ref{fig:NetworkProblems} and \ref{fig:NetworkProblemP01a} relating to Section 4.1 and Figures~\ref{fig:NetworkResults} and \ref{fig:NetworkResultR01d} relating to Section 4.2. In these figures, the height of boxes is a function of the grounded values associated with the domain/code it represents---the taller the box is, the better grounded the domain---, while the line width of the relationships is a function of the co-occurrence between domains/codes. From these networks, this section identifies and discusses the patterns between most relevant concepts at two levels of abstraction: domains and codes.

%%%%%%%%%%%%%%%%%%%%%%%%%%%%%%%%%%%%%%%%%%%%%%%%%%
\subsubsection{The problems that motivate a transition to DevOps}
From the networks of Figures~\ref{fig:NetworkProblems}  and \ref{fig:NetworkProblemP01a} we can induce some common behavior patterns as follows:

\begin{itemize}

\item Too much time for releasing (P01), digital transformation drivers (P10) (e.g., market trends, technological obsolescence), lack of standardization and automation (P09), and problems when releasing new versions (P02) are the most important concerns when adopting DevOps for participating organizations.

\item Too much time for releasing (P01), problems when releasing a new version (P02), and lack of standardization and automated processes (P09) may drive organizations to make digital transformations (P10), such as initiating large organizational changes or software architectural changes and modernizing technological stacks, many of which fall under the DevOps umbrella.

\item %The relationships between the domains P01, P02, and P10 (Figure~\ref{fig:NetworkProblems}) show that 
Due to market demands and trends (P10b), organizations may be forced to shorten releasing cycles; i.e., organizations need to be more agile and fast (P01a). In addition, participating organizations revealed the need to improve the quality of released products (P02a) so that the effort invested in high-quality software products and services may delay the release of new features. This means that the organizations participating in the study were not prepared to address the dilemma between speed and quality before DevOps adoption. %This can be also checked at the code level through the relationships between P01a, P02a, and P10b  (Figure~\ref{fig:NetworkProblemP01a}). 
%Hence, market needs (P10b) implies that organizations need to be more agile and fast (P01a), but at the same time they are forced to release new high quality products (P02a). This pattern is not new, but it is novel to see that organizations really think that DevOps is a solution to this dilemma.

\item The excessive time spent for releasing (P01) may be caused by problems when releasing (P02) (e.g., operations people are not able to deploy and release software because operational requirements were not considered during development), a lack of collaboration between dev \& ops (P07), organizational and cultural silos (P06), lack of standardization and automation (P09), and too much time spent on setting up environments (P03).

%\item The great number of problems associated with P01a \textit{``Need of being more agile, rapid''} is significant (Figure~\ref{fig:NetworkProblemP01a}). This is an indication that organizations think that DevOps is a possible solution to (i) reduce time-to-market, and (ii) adapt quickly to business needs and market demands. 

\item %The relationships between the domains P01, P06, and P07 (Figure~\ref{fig:NetworkProblems}) 
The data show that the lack of collaboration between development and operations teams generates silos, both cultural and organizational silos, and vice versa, the silos generate lack of collaboration. What seems clear is that \textit{\textbf{organizations identify the existence of silos and the lack of collaboration as a cause of the excessive time spent releasing new features}}. %(Figure~\ref{fig:NetworkProblems}, relationships P06-P01 and P07-P01).
In this regard, participating organizations revealed the need for a cultural change. 

\item The data show that organizational and cultural silos (P06) and the lack of collaboration (P07) make it difficult for organizations to standardize methods, tools, and technologies and automate processes (P09), which may be the main cause of problems when releasing new versions (P02) and may be the main cause of the excessive time spent setting up environments (P03).  

\item Barriers to innovation and experimentation (P05) have a low relevance (grounded = 2) and are not mentioned with or related to other domains (see Tables~\ref{tab:CodebookProblems} and \ref{tab:CoocurrencesProblemsDomain}, respectively). The fact that organizations have not brought up issues such as the flexibility and autonomy of the teams to enable and promote innovation and experimentation, as other studies (e.g., DASA reports) mention, is relevant. This means that either (i) organizations do not experience barriers to innovation, or (ii) organizations experience these barriers, but they do not associate this problem with the reasons that motivate them to adopt a DevOps culture. Our impression after analyzing all the data is that organizations initiated a DevOps transformation with the immediate objective of dealing with the excessive time spent on releasing features to satisfy their customers---which is their most important problem---but the capacity of rapid innovation and experimentation is a great goal to be dealt with in a second wave.           

\item The lack of end-to-end vision of value stream (P08) also has a low relevance  (grounded = 2) and is not mentioned with or related to other problems (see Tables~\ref{tab:CodebookProblems} and \ref{tab:CoocurrencesProblemsDomain}, respectively). Developers have too little focus on production environments and how systems work in these environments, while operations are involved in the development too late. The problems about the developers' lack of responsibility after they have put systems into production---non-shared end-to-end responsibility---, and the lower level of ownership that operations personnel possess with respect to developers were also revealed, although with less emphasis. 

\item System downtime (P04) is also rarely mentioned and rarely associated with other problems (see Tables~\ref{tab:CodebookProblems} and \ref{tab:CoocurrencesProblemsDomain}, respectively). This could be because most organizations participating in the study have values for mean time to recover (MTTR) that are suitable for their respective business domains.

\end{itemize}

%%%%%%%%%%%%%%%%%%%%%%%%%%%%%%%%%%%%%%%%%%%%%%%%%%
\subsubsection{The expected results with DevOps adoption}
From the networks of Figures~\ref{fig:NetworkResults}  and \ref{fig:NetworkResultR01d}, we can induce some common behavior patterns, as follows:

\begin{itemize}

\item Faster time-to-market (R01), better software quality (R02), process productivity (R03), and team effectiveness \& satisfaction (R04) were the most important results that the participating organizations expected to achieve when starting a DevOps adoption.

\item %The relationships between the domains R01, R03, and R04 (Figure~\ref{fig:NetworkResults}) 
The data show that the optimization and automation of processes (R03) may cause faster time-to-market (R01). Time-to-market may also be accelerated by increasing people's satisfaction and team effectiveness (R04). Thus, it seems that \textit{\textbf{organizations expect to accelerate time-to-market by improving processes and team efficiency, which drives the DevOps adoption}}. %This hypothesis is confirmed by the relationships between the codes R03b, R03a, R04c, R04d and the code R01d (Figure~\ref{fig:NetworkResultR01d}).

\item %The relationships between the domains R02, R03 and R04 (Figure~\ref{fig:NetworkResults}) 
Analogously, the data show that the optimization and automation of processes (R03) may improve software quality (R02). Software quality improvement may increase people's satisfaction and team effectiveness (R04). Thus, it seems that \textit{\textbf{organizations expect to improve software quality by improving process and team efficiency, which drives the DevOps adoption}}.

\item The data show that processes, people \& teams, and customers are intrinsically related. The participating organizations revealed that the optimization and automation of processes (R03) and team effectiveness and people's satisfaction (R04) are related to the improvement in customer satisfaction and experience (R05).

\item %The association between the domains R01 and R02 (Figure~\ref{fig:NetworkResults}) 
The data show that \textit{\textbf{organizations expect that the goal of reducing time-to-market will not be achieved at the expense of product quality}}. Specifically, the relationships of R01d with R02a (better software quality) and R02b (reliability, resilience, availability) may confirm this hypothesis.

\end{itemize}

%%%%%%%%%%%%%%%%%%%%%%%%%%%%%%%%%%%%%%%%%%%%%%%%%%
\subsubsection{About problems and results}

%Figure~\ref{fig:figure6} shows that, as a general rule, organizations identify the problems they have and that motivate them to adopt DevOps better than the expected results they can achieve. This demonstrates the general lack of knowledge about the promising benefits of DevOps culture, maybe motivated by the lack of empirical evidence.

%\begin{figure}[ht]
%\centering
%\includegraphics [width=12cm]{figure6.png}
%\caption{Relation between the identification of problems and expected results by organizations (in words)}
%\label{fig:figure6}
%\end{figure}

\begin{figure}[!h]
\centering
\includegraphics [width=12cm]{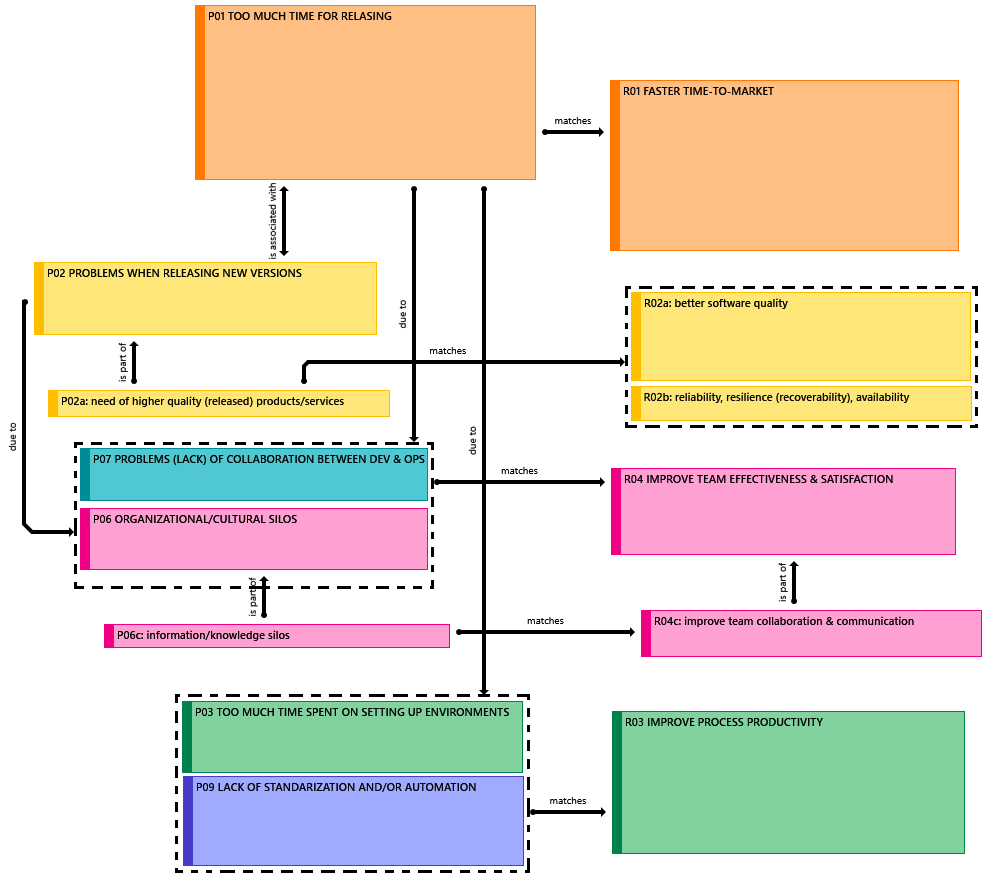}
\caption{Semantic network of problems and expected results}
\label{fig:NetworkProblemsResults}
\end{figure}

\begin{figure}[ht]
\centering
\includegraphics [width=9cm]{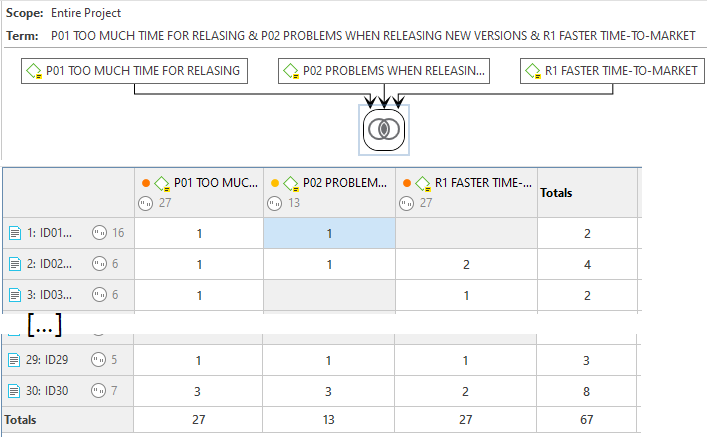}
\caption{Code-Document Table for P01, P02 and R01 by Atlas.ti}
\label{fig:figure7}
\end{figure}

%These relations among the above-mentioned problems and results are shown in the semantic network depicted in Figure~\ref{fig:NetworkProblemsResults}. The network establishes four types of relations: \textit{is associated with}, \textit{is cause of}, \textit{is part of}, and \textit{is addressed by}. %, which links a problem with the results that addressed it. %Finally, as in the previous networks, the size of the boxes (specifically the value of the height) indicates their grounded value, i.e. the higher the grounded, the taller the box.
Figure~\ref{fig:NetworkProblemsResults} depicts a semantic network that shows the existing relationships between the problems that lead an organization to adopt DevOps and the results they expect to obtain from such adoption. Semantic domains related to problems are on the left, and semantic domains related to results are on the right. The network establishes four types of relations: \textit{is associated with}, \textit{due to}, \textit{is part of}, and \textit{matches}. This network is described as follows:

\begin{itemize}

\item The expected result R01 \textit{``Faster time-to-market''} is a mirror of the problem P01 \textit{``Too much time for releasing''}, which in turn is associated with  P02 \textit{``Problems when releasing new versions''}, as is shown in Figure~\ref{fig:figure7} by querying data in Atlas.ti. Figure~\ref{fig:figure7} shows the code-document table for query P01 AND P02 AND R01. Although the table does not show the 30 cases, the resulting co-occurrence is 9.

\item The excessive time spent for releasing may be due to the problems P06 \textit{``Organizational and cultural silos''} and P07 \textit{``Lack of collaboration between Dev \& Ops''}, and the expected results when these problems are addressed are those embodied in R04 \textit{``Improve team effectiveness and satisfaction''}.

\item The excessive time spent for releasing may also be due to the problems P03 \textit{``Too much time spent on setting up environments''} and P09 \textit{``Lack of standardization and automation''}, and the expected results when these problems are addressed are those embodied in R03 \textit{``Improve process productivity''}.

\item Problems when releasing new versions may be mainly due to the problems P06 \textit{``Organizational and cultural silos''} and P07 \textit{``Lack of collaboration between Dev \& Ops''}. The organizational and cultural silos, and specifically the problem described by the code P06c \textit{``Information and knowledge silos''}, match R04c \textit{``Improve team collaboration $\&$ communication''}.

\item The organizations revealed the need for higher quality products and services (P02a), which  fits with the expected results of DevOps adoption: R02a \textit{``Better software quality''} and R02b \textit{``Reliability, resilience, availability''}.

\end{itemize}

%%%%%%%%%%%%%%%%%%%%%%%%%%%%%%%%%%%%%%%%%%%%%%%%%%
\subsubsection{DevOps Adoption Patterns and Anti-Patterns}

\begin{figure}[ht]
\centering
\includegraphics [width=12cm]{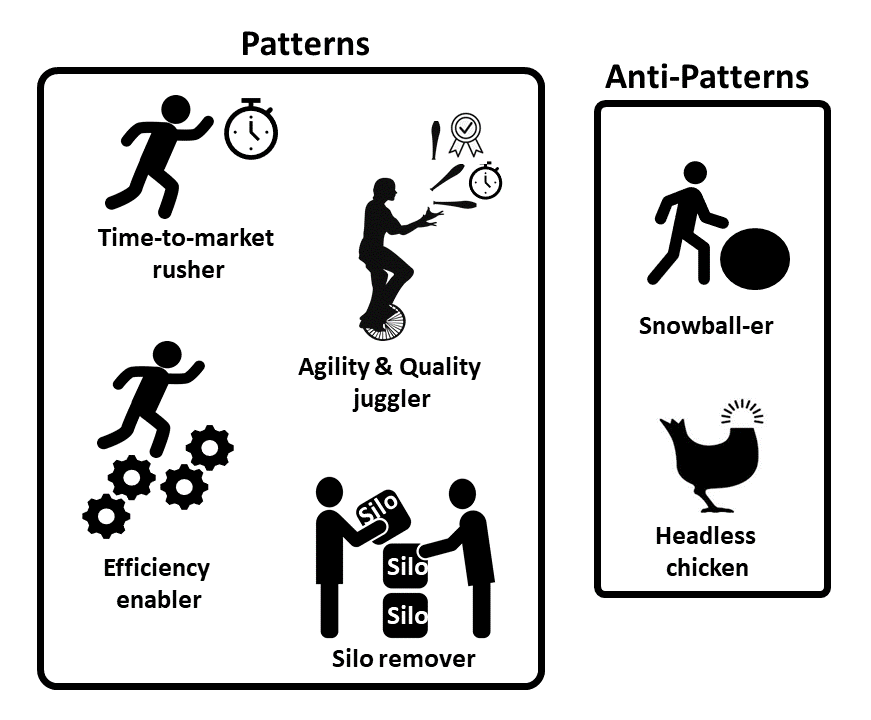}
\caption{Patterns and anti-patterns about the reasons that motivate companies to adopt DevOps}
\label{fig:patterns}
\end{figure}

The previous discussion about the problems that motivate companies to adopt DevOps and the expected results of this adoption allow us to discover several patterns and anti-patterns through a synthesizing process. These patterns and anti-patterns show the most relevant problems that trigger a DevOps transition and/or the most common expected results (see Figure~\ref{fig:patterns}); it is possible for an organization to manifest more than one pattern.

The patterns we identified are described as follows, and Table~\ref{tab:Patterns} shows the key points of any software pattern: intent, problem, solution, and benefits \cite{GoF}:

\begin{itemize}

\item  \textbf{Time-to-market rusher.} Reducing time to market and responding to market demands, either for new features or updates, are the two main drivers that lead companies to instill a DevOps culture. Organizations that start a DevOps transformation expect, above all, to reduce this time-to-market.

\item \textbf{Agility \& Quality juggler.} The market forces organizations to be more agile and faster, but at the same time, they are forced to release high-quality software products. Organizations that start a DevOps transformation expect to reduce time-to-market without sacrificing product quality. Indeed, they also expect to improve quality by introducing DevOps.

\item  \textbf{Silo remover.} The existence of organizational and cultural silos (e.g., matrix structures) causes a lack of collaboration between the development and operations departments, and in turn, this lack of collaboration generates knowledge silos. The existence of these silos and the lack of collaboration cause delays in releasing new versions. Organizations that start a DevOps transformation assume that the adoption of DevOps will improve the effectiveness and efficiency of (i) teams, i.e., by facilitating better communication and collaboration and by making teams more motivated, (ii) the process, due to the automation of continuous integration and deployment pipelines, and (iii) project management.

\item \textbf{Efficiency enabler.} Organizations that start a DevOps transformation expect that if they improve the efficiency of teams and processes (e.g., automation, standardization of technology stacks), they will improve the time-to-market and the product quality.

\end{itemize}

\begin{table}[!ht]
\centering
\small
\caption{DevOps adoption patterns}
\label{tab:Patterns}
\begin{tabular}{p{1.5cm} p{2.5cm} p{3cm} p{3cm} p{3.5cm}}
& \\
\hline
\textbf{Pattern} & \textbf{Intent} & \textbf{Problem} & \textbf{Solution} & \textbf{Benefits}\\ \hline 
Time-to-market rusher & Reduce time-to-market & Customer/market requires more velocity & Adopt DevOps focusing mainly on optimizing activities (e.g., automation) & Increase in customer satisfaction and business competitiveness by improving velocity\\ 

Agility \& Quality juggler & Increase quality, without harming velocity & Bugs are frequently reported, and hotfixes are usually needed. Overall, product quality could be improved & Adopt DevOps focusing mainly on quality (e.g., automatic testing) & Increase in customer satisfaction and business competitiveness by improving product quality\\

Silo \hspace{2.5cm} remover & Improve communication and collaboration between dev and ops & Existence of organizational, cultural and knowledge silos between Dev and Ops & Create DevOps teams focusing on the integration of people from dev and ops & Increase in synergies and team effectiveness \\

Efficiency enabler & Catch up with latest technologies and methodologies & The business is out of date at the technological and methodological levels & Adopt DevOps focusing on the incorporation of current technologies (e.g., CI/CD, micro-services) & Increase in business competitiveness by obtaining DevOps benefits\\
\hline
\end{tabular}
\end{table}

The anti-patterns we identified are described as follows, and Table~\ref{tab:Antipatterns} shows the key points of the software anti-patterns: intent, solution, emerging problem, and refactored solution \cite{Stamelos}.

\begin{itemize}
\item \textbf{Snowball-er.} Some organizations start a DevOps transformation as a consequence of a large architectural  (e.g., microservices) or infrastructural (e.g., cloud) change. Since they have to change, they also adopt DevOps, usually understanding DevOps as a process automation and ignoring any cultural and organizational approach. 

\item \textbf{Headless chicken.} Some organizations start a DevOps transformation because it is a trend or hype, or in the case of consulting organizations, because their clients ask for it, but they do not analyze the costs, benefits, and risks. 
\end{itemize}

\begin{table}[!ht]
\centering
\small
\caption{DevOps adoption anti-patterns}
\label{tab:Antipatterns}
\begin{tabular}{p{2.2cm} p{2.7cm} p{2.7cm} p{2.7cm} p{3.2cm}}
& \\
\hline
\textbf{Anti-pattern} & \textbf{Intent} & \textbf{Solution} & \textbf{Emerging problem} & \textbf{Refactored \hspace{1cm} solution} \\ \hline 

Snowball-er & Adopt DevOps as a consequence of other changes & Adopt too many technologies in many projects at once & The organization is not able to incorporate all the changes & Phasing-in the targeted changes \\

Headless chicken & Adopt DevOps to follow the latest trend, not knowing well what it is & Adopt inadequate technologies and methodologies & Unexpected and undesirable results & Thoughtful planning of the DevOps adoption process, sometimes with consultancy support \\

\hline
\end{tabular}
\end{table}

\section{Validity and Limitations of the Study}
\label{validity}
Case studies may have a number of threats concerning their validity. To discuss the possible threats, we have followed the four perspectives of validity threats presented by Wohlin et al. \cite{Wohlin:2012}: construct validity, internal validity, external validity and reliability.

\textit{Construct validity} concerns the procedure for collecting data and obtaining the right measures for the concept or the phenomenon being studied. %It addresses among others misinterpretation of interview questions, which was mitigated by discussing the interpretation of questions with the interviewees when was necessary. 
It is to an extent threatened by the ability of the interviewees to answer some questions that are beyond their scope or knowledge and to recall data. To mitigate this threat, we sent the interviewees, prior to the interview, the topics to be addressed. We interviewed key stakeholders, such as CEOs, CIOs, DevOps Platform leaders, product leaders, developers, and infrastructure managers (see Table \ref{tab:Interviewees}). These interviewees had all the necessary information to adequately answer the questions posed. Moreover, in some cases we contacted the interviewees several times until all questions or misinterpretations were addressed. Although the main data source of this study is interviews, we organized industrial workshops with some of the participating organizations\footnote{\url{http://bit.ly/2ky00LQ}, last accessed 2020/01/01.} and recorded observations from our visits to the organizations' premises in a \textit{research diary}. The data from workshops and observations were used to triangulate and complete data omissions. Finally, the transcriptions of interviews were sent to interviewees for inspection and approval. Therefore, prolonged contact with participants, data triangulation, and subsequent communication and approval, as our validity procedure established (see Section~\ref{method-ica}), allowed us to understand the interviewees' perspectives quite well. However, it is hard to observe a fully shared understanding of DevOps culture between practitioners and researchers; hence, some misinterpretation may remain, which is a limitation of the study.

\textit{Internal validity} concerns the identification of confounding factors that may have an effect on the outcomes. According to Yin \cite{Yin:2018}, internal validity is more relevant for explanatory and causal studies than for descriptive or exploratory studies. No strong causal claims are made due to the descriptive nature of this study, but the implied validity threat is acknowledged. Even so, we retrieved a rich description of the context of the involved organizations to identify possible confounding factors, such as the domain and size of the organizations, as our validity procedure established (see Section~\ref{method-ica}).

\textit{External validity} concerns the extent to which it is possible to generalize the findings, and to what extent the findings are of interest to other people outside the investigated case \cite{Wohlin:2012}. In case studies, there is no population from which a statistically representative sample can be drawn. We cannot claim full generalizability; our work is limited by the number of interviews, and the context of the participating companies. However, for case studies, the intention is to enable \textit{analytical generalization}, where the results are extended to cases that have common characteristics and, hence, for which the findings are relevant \cite{Wohlin:2012}. The more case studies are analyzed, the more likely companies that share context with participating companies can find relevant results. This can help mitigate some of the bias and validity threats inherent in qualitative research. We reached saturation with 30 companies, when the last few participants provided more evidence and examples but no new concepts or codes (we could have included fewer companies, but we added companies until we could be sure that new participants would not add new knowledge), as our validity procedure established (see Section~\ref{method-ica}).

\textit{Reliability} concerns the extent to which the data and the analysis are dependent on the specific researchers \cite{Wohlin:2012}. Ideally, a study should be able to be repeated by other researchers and achieve the same results \cite{Yin:2018}. In software engineering it is not usual to address this concern \cite{Cruzes:2011}, and authors conducting thematic analysis and other coding-based methods often state that they cannot guarantee that the coding of data in thematic analysis is entirely void of bias, as for example \cite{MAKINEN:2016}. However, in other knowledge areas in which qualitative analysis is much more advanced (such as psychology and sociology), reliability and, specifically, author bias is addressed using inter-coder agreement (ICA) techniques. In this study, we made a great effort to address this threat to validity, as our validity procedure established (see Section~\ref{method-ica}).

The next subsections describe the ICA analysis to assess the reliability and replicability of our study. For this purpose, we report the values of several Krippendoff's $\alpha$ coefficients achieved during the codification process. A precise definition and interpretation of these measures are summarized in Section~\ref{method-ica} and Appendix~\ref{appendix-ica}.

%%%%%%%%%%%%%%%%%%%%%%%%%%%%%%%%%%%%%%%%%%%%%%%%%%%%%%%%%%%%%%%%%%%%%%%%%%%%%%%%%%%
\subsection{Results for $\alphabin$}
\label{sec:val-alphabin}

In our case of study, the results for the $\alphabin$ coefficient for the research question RQ1 (problems) on the semantic domains defined in Section \ref{sec:RQ1}, for the two performed iterations, are shown in Table \ref{tab:table-alphabin}. As mentioned above, these results have been computed using the Atlas.ti software \cite{Atlas:2019}.

\begin{table}[h]
    \begin{center}
    \small
    \caption{Values of $\alphabin$ by semantic domain for each of the two evaluation iterations of the study for RQ1 (problems).}
    \begin{tabular}{c c c }
    \hline
        \textbf{Semantic domain} & \textbf{Iteration 1} & \textbf{Iteration 2} \\
    \hline
         P01 & 1.0 & 0.849 \\
    
         P02 & 0.651 & 0.655\\
    
         P03 & 1.0 & 1.0 \\
    
         P04 & 1.0 & 1.0\\
    
         P05 & 1.0 & 1.0\\
    
         P06 & 0.848 & 1.0\\
    
         P07 & 0.913 & -0.011\\
    
         P08 & 1.0 & 1.0\\
    
         P09 & 0.872 & 0.96\\
    
         P10 & 0.796 & 1.0 \\
    \hline
    \end{tabular}
    \label{tab:table-alphabin}
    \end{center}
\end{table}

From the results of Table~\ref{tab:table-alphabin}, we observe that the values of $\alphabin$ are generally high, providing evidences of agreement between the judges---i.e., Coder 1 and Coder 2 as described in Subsection~\ref{dataanalysis}---in the application of the different semantic domains. For iteration 1, the value of $\alphabin$ in 9 out of 10 of the semantic domains is above the minimal threshold $0.667$. Moreover, in 8 out of 10 of the semantic domains, the value is above $0.80$, showing statistical evidence of reliability in the evaluations. In iteration 2, the values of $\alphabin$ increased or remained equal in 8 out of 10 of the semantic domains. In this way, 8 out of 10 of the semantic domains present $\alphabin$ values above $0.667$ in iteration 2, and indeed, all of them are above $0.8$. However, we also observe two semantic domains that present more difficulties for being detected, namely, P02 and P07. 

In the case of P02, we observe that its $\alphabin$ values are consistently below the threshold of $0.667$ (but they increase from iteration 1 to iteration 2). This evidences indicates that although the agreement of the judges is clearly better than chance, the limits of this domain are not well defined. Indeed, this domain measures the analyzed companies’ needs for `higher quality' in several aspects of their development and operational cycle, which is intrinsically a very vague concept that opens the door to several interpretations from the judges. For this reason, to clarify the limits and applicability of this domain, future research is needed to propose quantitative metrics of quality. 

\begin{figure}[!ht]
\centering
\includegraphics [width=12cm]{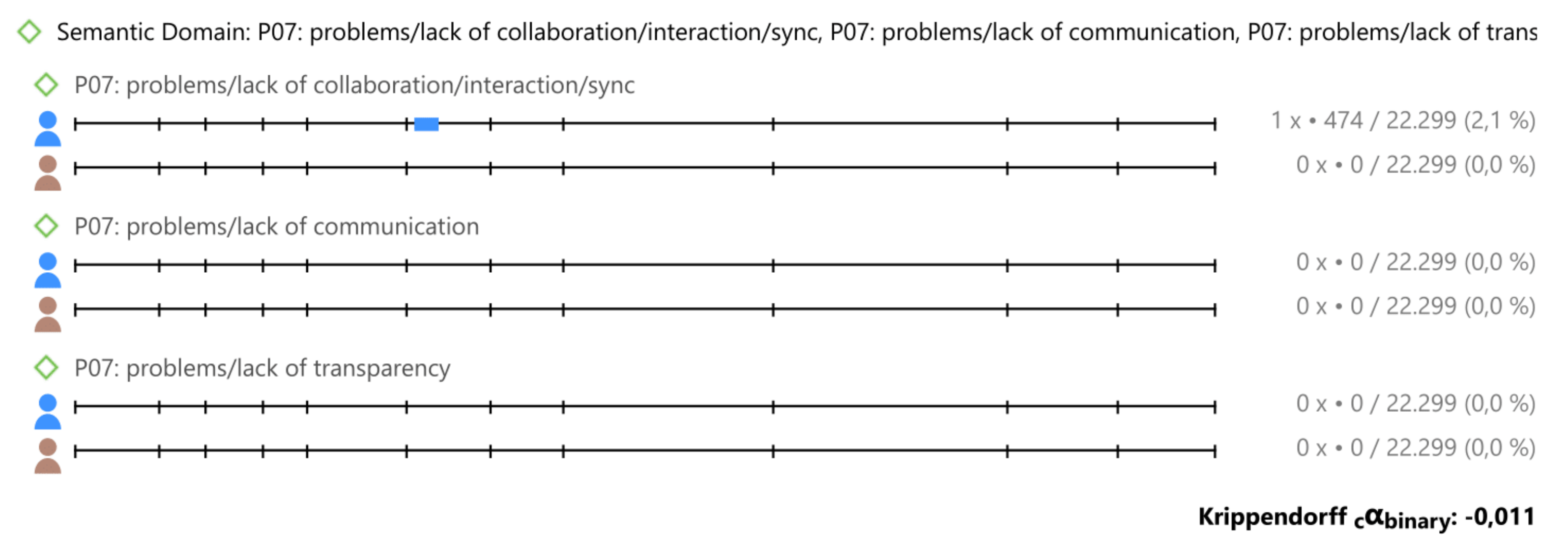}
\caption{Evaluation of semantic domain P07 during iteration 2 and its $\alphabin$ value.}
\label{fig:P07Binary}
\end{figure}

Finally, the case of P07 is special since the value dramatically decreases from iteration 1 to iteration 2 (where it is actually negative). The point is that, as shown in Figure \ref{fig:P07Binary}, in iteration 2, there is only one evaluation from one of the judges that assigns this semantic domain, in contrast with the 17 evaluations obtained in iteration 1. For this reason, there are not enough data for evaluating this domain in iteration 2, and thus, this result can be attributed to statistical outliers.

With respect to RQ2 (results), the obtained values of $\alphabin$ for the semantic domains described in Section \ref{sec:RQ2} are shown in Table \ref{tab:table-alphabin-result}. From these results, we observe that in iteration 1, all the semantic domains achieved $\alphabin$ values that were higher than the minimum threshold of significance, $0.667$. Furthermore, 5 out of 6 of the domains achieved values higher than $0.8$, showing statistical evidence of the reliability of the evaluations. Moreover, in iteration 2, the judges attained perfect agreement on the choice of the semantic domains, giving rise to a value of $\alphabin$ of $1.0$ in all the domains. This evidence indicates that the boundaries of these semantic domains are perfectly stated and that it is straightforward to determine whether to apply a code from a semantic domain.

\begin{table}[h]
    \begin{center}
    \small
    \caption{Values of $\alphabin$ by semantic domain on each of the two evaluation iterations of the study for RQ2 (results).}    
    \begin{tabular}{ c c c }
    \hline
        \textbf{Semantic domain} & \textbf{Iteration 1} & \textbf{Iteration 2} \\
    \hline
         R01 & 1.0 & 1.0 \\
    
         R02 & 0.763 & 1.0\\
    
         R03 & 0.822 & 1.0 \\
    
         R04 & 1.0 & 1.0\\
    
         R05 & 1.0 & 1.0\\
    
         R06 & 1.0 & 1.0\\
    \hline
    \end{tabular}

    \label{tab:table-alphabin-result}
    \end{center}
\end{table}

\begin{remark}
As mentioned in Subsection~\ref{dataanalysis}, in our study, the quotations were pre-established  before the evaluation of Coders 1 and 2. In this way, most of the content of the interviews was not selected as eligible since it was considered to be irrelevant for the analysis. This is consistent with the fact that relevant information in oral transmission is usually sparser than that in written media due to the cognitive needs of the speaker for elaborating a speech on the fly. However, a limitation of the software Atlas.ti \cite{Atlas:2019} forces us to consider the whole corpus in the ICA analysis and does not allow us to remove the irrelevant content from the analysis. This implies that the results are computed based on the whole corpus.
\end{remark}

%%%%%%%%%%%%%%%%%%%%%%%%%%%%%%%%%%%%%%%%%%%%%%%%%%%%%%%%%%%%%%%%%%%%%%%%%%%%%%%%%%%
\subsection{Results for $\cualpha$}
\label{sec:val-cu-alpha}

In our case of study, the results of $\cualpha$ for RQ1 (problems) are shown in Table \ref{tab:table-cualpha}. From these results, we observe that, in iteration 1, the value of $\cualpha$ for 9 out of 10 of the semantic domains is above the minimal threshold of $0.667$. Moreover, 7 out of 10 are above the confidence threshold of $0.8$. This shows that, regarding the scope of application of the codes within a particular domain, the results obtained in the study are sound and well founded. Indeed, 6 out of 10 of the coefficients are exactly $1.0$, meaning that there existed a perfect matching in the decisions of the judges. Moreover, in iteration 2, all the obtained coefficients consistently improved until reaching the value $1.0$; that is, there was perfect matching within all the domains.

\begin{table}[h]
    \begin{center}
    \small
    \caption{Values of $\cualpha$ by semantic domain for each of the two evaluation iterations of the study for RQ1 (problems).}    
    \begin{tabular}{c c c }
    \hline
        \textbf{Semantic domain} & \textbf{Iteration 1} & \textbf{Iteration 2} \\
    \hline
         P01 & 0.705 & 1.0 \\
    
         P02 & 1.0 & 1.0 (N/A)\\
    
         P03 & 0.962 & 1.0 \\
    
         P04 & 1.0 (N/A) & 1.0\\
    
         P05 & 1.0 & 1.0\\
    
         P06 & 0.739 & 1.0 (N/A)\\
    
         P07 & 1.0 & 1.0 (N/A)\\
    
         P08 & 1.0 & 1.0\\
    
         P09 & 1.0 & 1.0\\
    
         P10 & 0.563 & 1.0 \\
    \hline
    \end{tabular}

    \label{tab:table-cualpha}
    \end{center}
\end{table}

\begin{figure}[!ht]
\centering
\includegraphics [width=12cm]{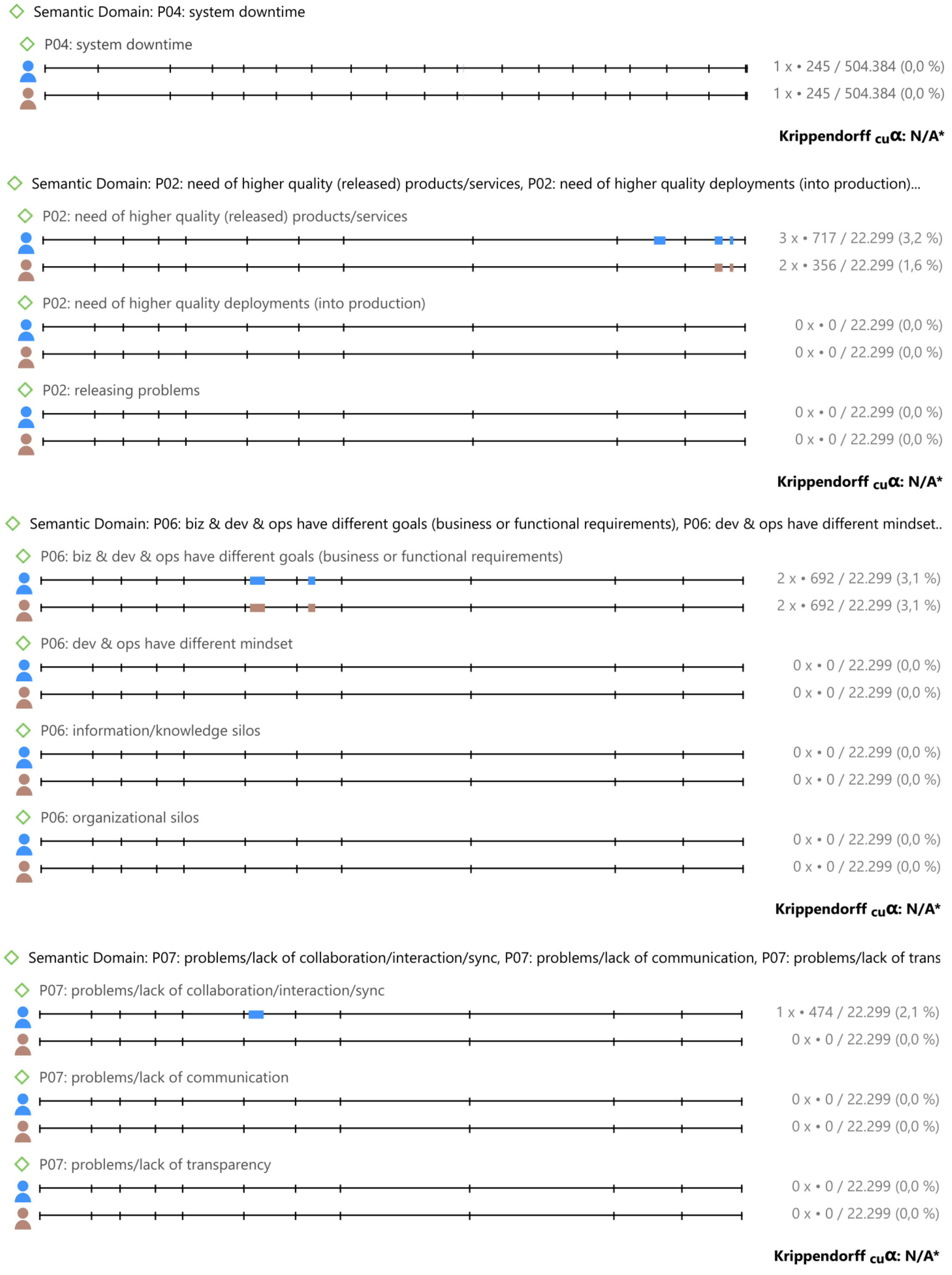}
\caption{Evaluation of semantic domains P04 (iteration 1) and P02, P06 and P07 (iteration 2) and their $\cualpha$ values.}
\label{fig:cuProblems}
\end{figure}

It is worth mentioning that, due to the limitations of the software used, some of the perfect matches were labeled N/A. The reason is that Atlas.ti considers that the results present insufficient variability for computing $\cualpha$. For instance, domain P04 in iteration 1 was only chosen once by the two judges. In the same vein, for domains P02, P06 and P07 of iteration 2, only one code was used in each. These claims can be checked in Figure \ref{fig:cuProblems}. For these reasons, the software displays that the results are not available, even though, strictly speaking, the $\cualpha$ coefficient is $1.0$, as stated in Table \ref{tab:table-cualpha}.

\begin{figure}[!ht]
\centering
\includegraphics [width=12cm]{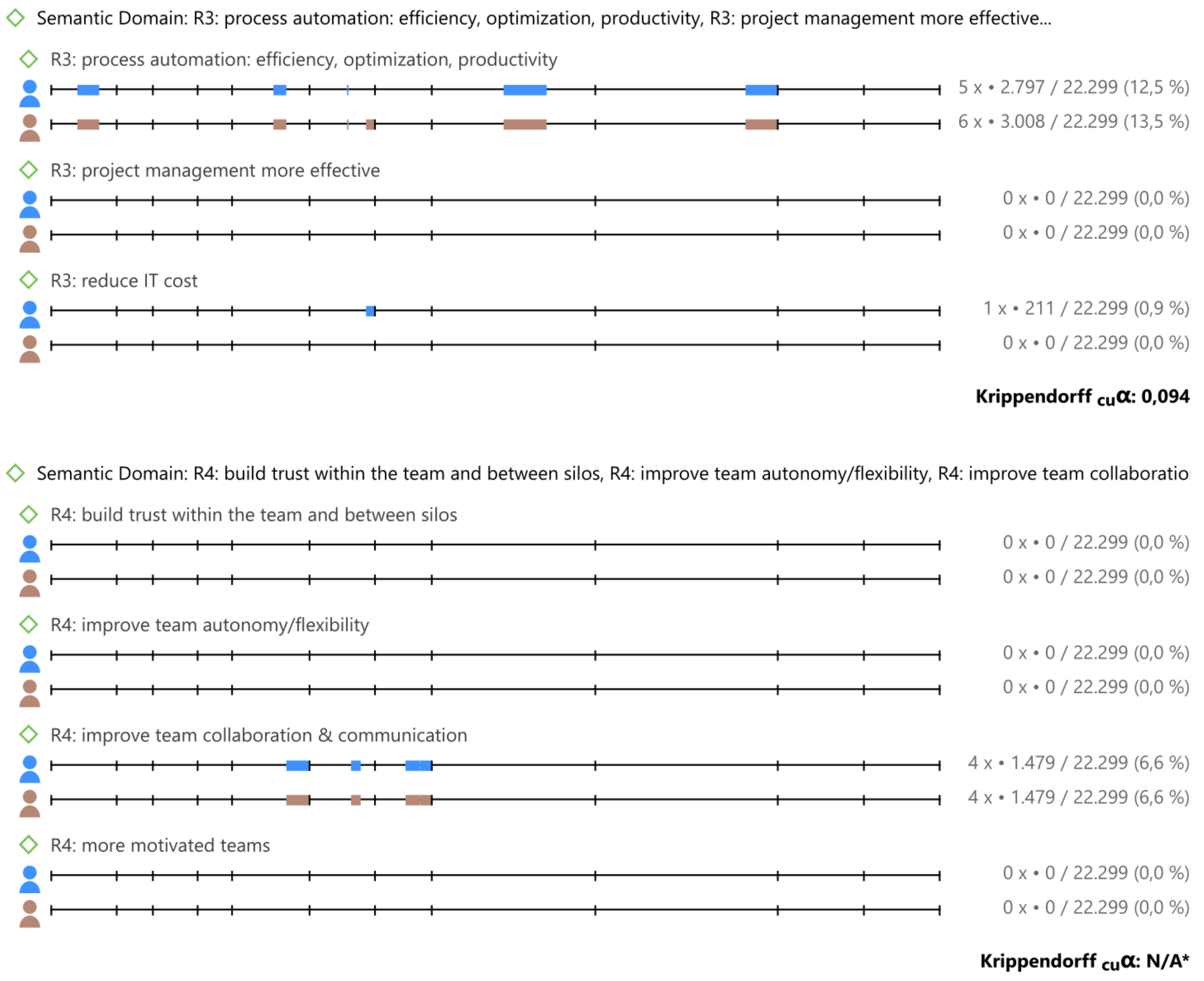}
\caption{Evaluation of semantic domains R03 and R04 during iteration 2 and their $\cualpha$ values.}
\label{fig:cuResults}
\end{figure}

On the other hand, the results of $\cualpha$ for RQ2 (results) are shown in Table \ref{tab:table-cualpha-result}. Overall, the values of $\cualpha$ for this research question are high. Indeed, in iteration 1, the judges obtained perfect matching in the choice of the codes within a semantic domain, implying a value of $\cualpha$ of $1.0$. In iteration 2, the results follow the same lines, but we find a couple of values that deserve an explanation. These are the values of $\cualpha$ for the semantic domains R03 and R04, whose evaluations can be checked in Figure \ref{fig:cuResults}. In the case of R04, as mentioned above, it was labelled N/A by Atlas.ti because the software considered that there was not enough statistical variability. In this particular case, the two judges picked only one code from the semantic domain, but they agreed on the choice, so this is, indeed, a perfect matching.

\begin{table}[h]
    \begin{center}
    \small
    \caption{Values of $\cualpha$ by semantic domain on each of the two evaluation iterations of the study for RQ2 (results).}    
    \begin{tabular}{c c c}
    \hline
        \textbf{Semantic domain} & \textbf{Iteration 1} & \textbf{Iteration 2} \\
    \hline
         R01 & 1.0 & 1.0 \\
    
         R02 & 1.0 & 1.0\\
    
         R03 & 1.0 & 0.094 \\
    
         R04 & 1.0 & 1.0 (N/A)\\
    
         R05 & 1.0 & 1.0\\
    
         R06 & 1.0 & 1.0\\
    \hline
    \end{tabular}

    \label{tab:table-cualpha-result}
    \end{center}
\end{table}

In the case of R03, the small value of $\cualpha$ is produced by the unbalanced use of its codes. First, of the three codes within R03, only two were used by the judges, namely R03a \textit{``Process automation: efficiency, optimization, productivity''} and R03c \textit{``Reduce IT cost''}. However, R03a is much more popular than R03c since it received all the evaluations from the coders except a short one.

In this way, the expected agreement, $A_e$, for R03 is very high, since with high probability  and regardless of the particular quotation, all the coders choose R03a and thus agree. This implies that the expected disagreement is artificially low, $D_e \approx 0$. This forces the quotient of the observed disagreement and the expected disagreement to be very high, $D_o/D_e \gg 0$, even though $D_o$ might be quite small (as in this case, with a disagreement in only 1 of the 6 quotations). This has the paradoxical consequence that, even if very few disagreements take place, the $\cualpha$ coefficient is very small, $\cualpha = 1- D_o / D_e \approx 0$. In our case, this  disagreement provokes a large distortion of the results. Therefore, this value is not statistically significant, and indeed, both judges achieved a high degree of agreement in this semantic domain, comparable with their results in iteration 1.

%However, the use of these two codes is deeply unbalanced: the former code received all the evaluations from the judges except a short quotation that was coded by one judge with the later code. This unbalanced situation produces that, through the eyes of the $\cualpha$, most of the coincidences are due to the chance, so the only disagreement provokes a large distortion of the result. In this way, this value is not statistically significant and, indeed, both judges achieved a high degree of agreement in this semantic domain, comparable with their results in iteration 1.

%%%%%%%%%%%%%%%%%%%%%%%%%%%%%%%%%%%%%%%%%%%%%%%%%%%%%%%%%%%%%%%%%%%%%%%%%%%%%%%%%%%
\subsection{Results for $\Cualpha$}

Table \ref{tab:table-Cualpha} shows the $\Cualpha$ values for iterations 1 and 2 and for RQ1 (problems) and RQ2 (results).

\begin{table}[h]
    \begin{center}
    \small
    \caption{Values of $\Cualpha$ on each of the two evaluation iterations of the study for the research questions RQ1 and RQ2.}    
    \begin{tabular}{c c c}
    \hline
        \textbf{$\Cualpha$} & RQ1 (problems) & RQ2 (results) \\
    \hline
         Iteration 1 & 0.67 & 0.911 \\
    
         Iteration 2 & 0.905 & 0.98 \\
    \hline
    \end{tabular}

    \label{tab:table-Cualpha}
    \end{center}
\end{table}

With respect to RQ1 (problems), in iteration 1, we obtained a global $\Cualpha$ coefficient of $0.67$, which is slightly above the lower threshold of applicability, $0.667$. This is compatible with the observation of Section 5.1 that, for this iteration, $\alphabin$ reached significantly low values for semantic domains P02 and P09. This is evidence that the boundaries of some semantic domains in the first version of the codebook were not well defined and their descriptions presented a certain amount of overlapping. Nevertheless, this problem was overcome with the new version of the codebook. In iteration 2, the value of $\Cualpha$ increased to $0.905$. This is clearly above the threshold of $0.8$ reported in the literature for obtaining statistical significance. This evidences that the review process of the codebook detected the flaws in the description of the limits of applicability of the semantic domains, which led to clearer and more bounded definitions.

With respect to RQ2 (results), we observed that $\Cualpha$ obtained very high values, higher than 0.8, in both iterations. This confirms the trend previously observed that the semantic domains of this research question are clearly stated. Indeed, the value of $\Cualpha$ increased in iteration 2 with respect to the value of iteration 1, in accordance with the previous observation for RQ1.

\section{Related Work}
\label{related}
Some systematic literature reviews, systematic mapping studies, and surveys on both gray and technical literature, have been previously conducted to answer RQ1 and RQ2 \cite{slr1,Erich:2019,Erich:2014,slr7,slr8,slr13}. Recently, Leite et al. \cite{Leite:2020} published an exhaustive survey on DevOps concepts and challenges, in which through a method inspired by systematic literature review and grounded theory, they analyze practical implications for engineers, managers and researchers. However, we focus on empirical evidence based on exploratory case studies and grounded theory studies in which units of analysis are companies adopting DevOps (i.e., we do not focus on literature reviews or studies on companies adopting specific practices, such as continuous delivery or microservice architectures). Experience reports or any other studies that are not based on empirical evidence from companies adopting DevOps are not considered in the following analysis. Chronologically, although the study by Iden et al. \cite{Iden:2011} did not explicitly mention DevOps, we can consider that is one of the first papers that empirically analyzed the conflict between development and operations teams when they have to collaborate. This follows the idea initiated by P. Debois \cite{Debois:2008} and Flickr employees \cite{Allspaw:2009}.  Iden et al. \cite{Iden:2011} used the Delphi method (brainstorming with 42 Norwegian IT experts, reduction and ranking) to provide a key baseline for analyzing the problems that reveal the lack of cooperation between developers and IT operations personnel. 

Some years later, 
%Erich  et al. \cite{Erich:2014} performed a systematic mapping study to analyze the benefits of this cooperation between development and operations, and 
Smeds et al. \cite{Smeds:2015} interviewed 13 subjects in a software company adopting DevOps to research the main defining characteristics of DevOps and the perceived impediments to DevOps adoption. Later, Lwakatare et al. \cite{Lwakatare:2016b} used multi-vocal literature and three interviews from one case company to describe what DevOps is and outline DevOps practices according to software practitioners. In the same year, Riungu-Kalliosaari et al. \cite{Lwakatare:2016a} conducted a qualitative multiple-case study and interviewed the representatives of three software development organizations in Finland to answer how industry practitioners perceive the benefits of DevOps practices in their organization and how they perceive the adoption challenges related to DevOps.

In the following years a greater number of studies conducted empirical research by involving an increasing number of companies. Erich et al. \cite{Erich:2017,Erich:2019} performed an exploratory study on six companies to answer again what DevOps is and to explore the effects when DevOps is practiced, how DevOps is implemented and what supportive factors exist. Kuusinen et al. \cite{Kuusinen:2018} also conducted a case study in a large Danish software company to identify challenges, impediments, and barriers that a large company faces when transitioning towards DevOps. Senapathi et al. \cite{Senapathi:2018} conducted an exploratory case study that explored DevOps implementation in a New Zealand product development organization to research the main drivers for adopting DevOps, the engineering capabilities and technological enablers of DevOps, and the benefits and challenges of using DevOps. This study, for the first time, considered why companies are instilling a DevOps culture, although only in a large organization in the finance/insurance sector. Finally, Luz et al. \cite{LuzPinto:2019} conducted grounded theory about 15 scenarios of successful DevOps adoption in companies to build a model. 

These academic studies, all of them based on empirical research, provide a key baseline for future studies with a broader scope until achieving saturation for qualitative studies (see Table~\ref{tab:empiricalstudies}). Table~\ref{tab:empiricalstudies} summarizes the companies involved in case studies and grounded theory studies conducted until the present, and Table~\ref{tab:relatedwork} shows the research questions addressed in these studies. These studies mainly focused on DevOps definition and characterization, practices, benefits, and challenges. However, the drivers and problems that lead companies to adopt DevOps, as well as the expectations of this cultural and organizational transformation, received little attention in these studies. 

\begin{table}[!h]
\centering
\small

\caption{List of empirical studies (\#C: number of companies; \#P: number of participants)}
\label{tab:empiricalstudies}
\begin{tabular}{p{7.5 cm} p{6cm} p{0.8cm} p{1.0cm}}
& \\
\hline
\textbf{Research methodology - collection method} & \textbf{Author(s) (year)} & \textbf{\#C}  & \textbf{\#P}\\ \hline 

 Delphi method & Iden et al. (2011) \cite{Iden:2011} & - & 42 \\
 Case study - semi-structured interviews & Smeds et al (2015) \cite{Smeds:2015} &  1 & 13 \\
 Case study - semi-structured interviews & Lwakatare et al. (2016) \cite{Lwakatare:2016b} & 1 & 3 \\
 Multiple-case study - semi-structured interviews & Riungu-Kalliosaari et al. (2016) \cite{Lwakatare:2016a} & 3 & 3\\
 Multiple-case study - semi-structured interviews & Erich et al. (2017) \cite{Erich:2017,Erich:2019} & 6 & 6\\
 Case study - survey + interviews & Kuusinen et al. (2018) \cite{Kuusinen:2018} & 1 & 34 + 4 \\
 Case study - semi-structured interviews & Senapathi et al. (2018) \cite{Senapathi:2018} & 1 & 6 \\
 Grounded Theory - semi-structured interviews & Pinheiro Luz et al. (2019) \cite{LuzPinto:2019} & 15 & - \\
 Grounded Theory - semi-structured interviews & Rafi et al. (2020) \cite{Saima:2020} & 5 & 13 \\
 Grounded Theory - semi-structured interviews & Leite et al. (2020) \cite{LeitePinto:2020} & $<27$ & 27 \\
\hline
\end{tabular}

\end{table}

\begin{table}[!h]
\centering
\small
\caption{Research questions addressed in related studies}
\label{tab:relatedwork}
\begin{tabular}{p{8cm} p{4cm}}
& \\
\hline
\textbf{Research Question} & \textbf{References}\\ \hline 
DevOps concept and characteristics &  \cite{Smeds:2015} \cite{Lwakatare:2016b}  \cite{Erich:2017} \\
DevOps drivers - problems & \cite{Senapathi:2018} \cite{LuzPinto:2019} \\
DevOps expectations - perception of benefits & \cite{LuzPinto:2019} \\
DevOps enablers - supportive factors &  \cite{Erich:2017} \cite{Senapathi:2018} \\
DevOps practices - how is DevOps implemented & \cite{Lwakatare:2016b} \cite{Erich:2017} \cite{dora:2018} \cite{puppet:2018}  \\
DevOps benefits - effects &  \cite{Erich:2014} \cite{Lwakatare:2016a} \cite{Erich:2017} \cite{Senapathi:2018} \cite{dora:2018} \\
DevOps challenges - impediments  & \cite{Smeds:2015} \cite{Lwakatare:2016a} \cite{Senapathi:2018} \cite{Kuusinen:2018}   \\
Security concerns in DevOps adoption & \cite{Saima:2020} \\
Team Topologies &  \cite{LeitePinto:2020}\\
\hline
\end{tabular}
\end{table}

Erich et al. \cite{Erich:2017} pointed out the need for more experimental studies and quantitative studies to verify the state of DevOps. In this regard, both the report made by DORA (DevOps Research \& Assessment association) \cite{dora:2018} and the report made by Puppet and Splunk \cite{puppet:2018}, analyzed data from survey questionnaires distributed to over 30,000 technical professionals worldwide. The first one identifies a set of software delivery performance profiles (elite, high, medium and low performance) and relates DevOps practices with these profiles. The second one identifies five stages of DevOps evolution (aka. the DevOps evolutionary model) and establishes the practices that define and/or contribute to success in each stage. These reports also provide a valuable information for companies as they provide a global picture, but they do not respond to why many businesses are instilling a DevOps culture. 

Between these large  surveys and the abovementioned qualitative studies, we presented this multiple case study of 30 software-intensive companies, in which 44 relevant stakeholders participated to answer RQ1 and RQ2 and which satisfies the criteria of saturation for qualitative studies. Focusing on RQ1 and RQ2 the most similar study is that by Luz et al. \cite{LuzPinto:2019}, which analyzes 15 companies. This study and ours coincide in the basic constructors, and each one contributes with new constructors; therefore, they are quite complementary in the search to understand what motivates organizations to move to DevOps and what outcomes they expect to obtain. Luz et al. pointed out the importance of adopting a collaborative culture to remove silos between development and operations teams and this is consistent with our results (see Table~\ref{tab:CodebookProblems}). However, our study highlights that organizations identify the existence of silos and the lack of collaboration as a main cause of the excessive time spent  releasing new features, and this is the most relevant problem that moved companies to adopt DevOps. %, since many companies mentioned as problems that motivate them to incorporate DevOps the existence of organizational and cultural silos (P06, grounded=10) and problems or lack of collaboration between Dev \& Ops (P07, grounded=9). 
%We identify, as Luz et al. do, that collaboration is so critical because it is strongly related to other issues such as agility or automation (see Figure~\ref{fig:NetworkProblems}). On the other hand, Luz et. al also pointed out the importance of automation to support DevOps adoption. Once again, this is consistent with our results. The lack of standardization and/or automation (P09, grounded=14) and the excessive time in setting up environments (P03, grounded=9) was mentioned by many companies as problems that motivate them to adopt DevOps. 
Last, Luz et al. highlighted agility and resilience as expected outcomes of DevOps adoption, which are two of the most relevant results of adopting DevOps for the companies participating in our study (see Table~\ref{tab:CodebookResults}). However, our study also identified that organizations expect to improve process productivity and team effectiveness \& satisfaction, provide more value to customers, and align the objectives of the development and operations teams with business.  

Finally, none of the mentioned related work applies methods for evaluating the reliability and consistency when coding qualitative data. Generally, when thematic analysis or other coding-based analysis methods are used in qualitative research in software engineering, inter-coder reliability is not calculated; thus, author bias in findings may exist, and it remains undetected.

\section{Conclusion}
\label{conclusions}
This paper provides empirical evidence about the reasons why companies move to DevOps and what results they expect to obtain when adopting a DevOps culture. This paper describes empirical research on practicing DevOps through a multiple case study of 30 multinational software-intensive companies that mainly consists of interviews with 44 key stakeholders, together with observations and memos gathered in the research diary. To improve the validity of this exploratory case study we applied various strategies, such as data triangulation, member checking, rich description, clarification of bias, and discrepant reporting. The method used in this paper to reduce author bias, the inter-coder agreement based on some variants of Krippendorff's $\alpha$ coefficient, is especially relevant. All these strategies mitigate the problems inherent in qualitative research and reinforce our findings.

%About the study findings is relevant for us to highlight: 
The main problem that motivates software companies to adopt DevOps is that delivering software takes too much time. DevOps culture and practices promote higher levels of process automation and efficiency and team effectiveness and collaboration. This leads to faster time-to-market and  contributes to improving software quality and customer satisfaction.

We have discovered some patterns and anti-patterns about the reasons why companies are instilling a DevOps culture in their organization (see Figure~\ref{fig:patterns}); i.e., our results show the most relevant problems that trigger a DevOps transition and/or the most common expected results. 

Some patterns focus on a cultural and organizational perspective, emphasizing the silos and the lack of collaboration between dev \& ops and the need to break these silos and improve collaboration (\textit{silo remover}), while other patterns focus on team and process efficiency, emphasizing the lack of process automation and standardized technology stacks, infrastructure, methodologies, etc., and the need for optimizing and automating processes (\textit{efficiency enabler}). What is clear is that most organizations adopt DevOps with the goal of accelerating time-to-market (\textit{time-to-market rusher}) while delivering high-quality products (\textit{agility $\&$ quality juggler}).

Some anti-patterns regard the fact that some organizations adopt DevOps as a trend or hype (\textit{headless chicken} anti-pattern) and the fact that they adopt DevOps while simultaneously facing a large change (e.g. adopting a new reference architecture or a new infrastructure for core applications, such as OpenShift) likely incurring a \textit{Snowballing} anti-pattern.

As further work we plan to provide more empirical evidence about other research questions, such as, team topologies, key performance attributes, such as lead time, deployment frequency, and mean time to recovery, and other barriers to adopting DevOps. 

\section*{Acknowledgement}
This research project is being performed thanks to Vass, Clarive, Autentia, Ebury, Carrefour, Vilt, IBM, AtSistemas, Entelgy, Analyticalways, Mango eBusiness, Adidas, Seur, Zooplus, as well as other participating companies.

\appendix
\section{Appendix. Theoretical framework for $\alpha$ coefficients}
\label{appendix-ica}
In this Appendix, we introduce a novel interpretation that unifies the different variants of the $\alpha$ coefficient into a common framework. These coefficients that may be found in the literature are presented as unrelated and a kind of ad hoc formulation for each problem is provided.

The key point of this section is that we will show that these versions can be translated to simpler and universal version of Krippendoff's $\alpha$. For this purpose, we will formulate the $\alpha$ coefficients in terms of some `meta-codes', that we will call `labels'. In each situation, we will provide a well-defined algorithm that translates from semantic domains and codes (the units of judgment considered in thematic analysis) into labels. In this way, after this translation, all the coefficients reduce to the same mathematical computation of the universal $\alpha$ coefficient for labels.

In order to lighten this section, the mathematical formulation of this universal $\alpha$ coefficient has been moved to Appendix \ref{appendix-alpha}. Nevertheless, for convenience, let us recall some notations introduced there that will be used along this section. We fix a finite set $\Lambda$ of labels and we are dealing with a collection $J_1, \ldots, J_n$ of judges that will evaluate a set of items $I_1, \ldots, I_m$. Each of the judges, $J_\alpha$, evaluates the item $I_\beta$ with a subset $\omega_{\alpha, \beta} \subseteq \Lambda$. The result of the evaluation process is gathered in a set $\Omega = \left\{\omega_{\alpha, \beta}\right\}$ for $1 \leq \alpha \leq n$ and $1 \leq \beta \leq m$.

From $\Omega$, we can compute Krippendoff's coefficient, $\alpha = \alpha(\Omega)$, which is a real number with $0 \leq \alpha \leq 1$. Such a quantity is interpreted as a measure of the degree of agreement that is achieved out of the chance. The bigger the $\alpha$ is, the better agreement is observed. A common rule-of-thumb in the literature \cite{Krippendorff:2018} is that $\alpha \geq 0.667$ is the minimal threshold required for drawing conclusions from the data. For $\alpha \geq 0.80$, we can consider that there exists statistical evidence of reliability in the evaluations.

However, this ideal setting, as described in Appendix~\ref{appendix-alpha}, might be too restrictive for the purposes of content analysis (particularly, as applied by the Atlas.ti Software \cite{Atlas:2019}). The most general setting of content analysis is as follows. We have a collection of $s > 1$ semantic domains, $S_1, \ldots, S_s$. A semantic domain defines a space of distinct concepts that share common meanings (for a concrete example, check our semantic domains in Subsection~\ref{subsec:codebookproblems}). Subsequently, each semantic domain embraces mutually exclusive concepts indicated by a code. Hence, for $1 \leq i \leq s$, the domain $S_i$ for $1 \leq i \leq s$, decomposes into $r_i \geq 1$ codes, that we will denote by $C_1^i, \ldots, C_{r_i}^i$. As pointed out in the literature, for design consistency these semantic domains must be logically or conceptually independent. This principle translates into the fact that there exist no shared codes between different semantic domains.

Now, the data under analysis (e.g. scientific literature, newspapers, videos,  interviews) is divided into items, which in this context are known as \textit{quotations}, that represent meaningful parts of the data by their own. The decomposition may be decided by each of the judges (so different judges may have different quotations) or it may be pre-established (for instance, by the codebook creator or the designer of the ICA study). In the later case, all the judges share the same quotations so they cannot modify the limits of the quotations and they should evaluate each quotation as a block. To enlighten the notation, we will suppose that we are dealing with this case of pre-established quotations. This is the setting of our study (see Subsection~\ref{dataanalysis}). Indeed, from a mathematical point of view, the former case can be reduced to this version by refining the data division of each judge to get a common decomposition into the same pieces.

Therefore, we will suppose that the data is previously decomposed into $m \geq 1$ items or quotations, $I_1, \ldots, I_m$. Observe that the union of all the quotations must be the whole matter so, in particular, irrelevant matter is also included as quotations. Now, each of the judges $J_\alpha$, $1 \leq \alpha \leq n$, evaluates the quotation $I_i$, $1 \leq i \leq m$, assigning to $I_i$ any number of semantic domains and, for each chosen semantic domain, one and only one code. No semantic domain may be assigned in the case that the judge considers that $I_\beta$ is irrelevant matter, and several domains can be applied to $I_\beta$ by the same judge.

Hence, as byproduct of the evaluation process, we obtain a collection of sets $\Sigma = \left\{\sigma_{\alpha, \beta}\right\}$, for $1 \leq \alpha \leq n$ and $1 \leq \beta \leq m$. Here, $\sigma_{\alpha, \beta} = \left\{C_{j_1}^{i_1}, \ldots, C_{j_p}^{i_p}\right\}$ is the collection of codes that the judge $J_\alpha$ assigned to the quotation $I_\beta$. The exclusion principle of the codes within the semantic domain means that the collection of chosen semantic domains $i_1, \ldots, i_p$ contains no repetitions.

\begin{remark}
To be precise, as proposed in \cite{Krippendorff:2016}, when dealing with a continuum of matter each of the quotations must be weighted by its length in the observed and expected coincidences matrices (see Appendix \ref{appendix-alpha}). This length is defined as the amount of atomic units the quotation has (say characters in a text or seconds in a video). In this way, (dis)agreements in long quotations are more significant than (dis)agreements in short quotations. This can be easily incorporated to our setting just by refining the data decomposition to the level of units. In this way, we create new quotations having the length of an atomic unit. Each new atomic quotation is judged with the same evaluations as the old bigger quotation. In the coefficients introduced below, this idea has the mathematical effect that, in the sums of Equation (\ref{eq:disagreement}, Appendix~\ref{appendix-alpha}), each old quotation appears as many times as atomic units it contains, which is the length of such quotation. Therefore, in this manner, the version explained here computes the same coefficient as in \cite{Krippendorff:2016}.
\end{remark}

In order to quantify the degree of agreement achieved by the judges in the evaluations $\Sigma$, several variants of Krippendorff's $\alpha$ are proposed in the literature \cite{Krippendorff:2016,Krippendorff:2018}. For the purposes of this study, we will apply the variants described below.

%%%%%%%%%%%%%%%%%%%%%%%%%%%%%%%%%%%%%%%%%%%%%%%%%%%%%%%%%%%%%%%%%%%%%
\subsubsection{The coefficient $\alphabin$}
\label{sec:alphabin}

The first variation of the Krippendorff's $\alpha$ coefficient is the so-called $\alphabin$ coefficient. This is a coefficient that must be computed on a specific semantic domain. Hence, let us fix a semantic domain $S_{i}$ for some fixed $i$ with $1 \leq i \leq s$. The set of considered items to be judged is exactly the set of (prescribed) quotations $I_1, \ldots, I_m$. However, the set of labels will have only two labels, that semantically represent `voted $S_{i}$' and `did not vote $S_{i}$'. Hence, we take
$$
    \Lambda = \left\{1, 0\right\}.
$$

For the assignment of labels to items, the rule is as follows. For $1 \leq \alpha \leq n$ and $1 \leq \beta \leq m$, we set $\omega_{\alpha,\beta} = \left\{1\right\}$ if the judge $J_\alpha$ assigned some code of $S_{i}$ to the quotation $I_\beta$ (i.e.\ if $C_j^{i} \in \sigma_{\alpha, \beta}$ for some $1 \leq j \leq r_{i}$) and $\omega_{\alpha,\beta} = \left\{0\right\}$ otherwise. Observe that, in particular, $\omega_{\alpha,\beta} = \left\{0\right\}$ if $J_\alpha$ considered that $I_\beta$ was irrelevant matter. From this set of evaluations, $\Omega_{binary}^{S_{i}} = \left\{\omega_{\alpha, \beta}\right\}$, $\alphabin$ is given as
$$
    \alphabin^{S_{i}} = \alpha(\Omega_{binary}^{S_{i}}).
$$

In this way, the coefficient $\alphabin^{S_{i}}$ can be a measure of the degree of agreement that the judges achieved when choosing to apply the semantic domain $S_{i}$ or not. A high value of $\alphabin^{S_{i}}$ is interpreted as an evidence that the domain $S_{i}$ is clearly stated, its boundaries are well-defined and, thus, the decision of applying it or not is near to be deterministic. However, observe that it does not measure the degree of agreement in the application of the different codes within the domain $S_{i}$. Hence, it may occur that the boundaries of the domain $S_{i}$ are clearly defined but the inner codes are not well chosen. This is not a task of the $\alphabin^{S_{i}}$ coefficient, but of the $\cualpha^{S_{i}}$ coefficient explained below.

\begin{remark}
Empirically, we discovered that the semantic that the software Atlas.ti \cite{Atlas:2019} applies for computing $\alphabin$ (and for the coefficient $\cualpha$ introduced in Section \ref{sec:cu-alpha}) is the one explained in this section. However, to our understanding, this behavior is not consistent with the description provided in the corresponding user's guide.
\end{remark}

%%%%%%%%%%%%%%%%%%%%%%%%%%%%%%%%%%%%%%%%%%%%%%%%%%%%%%%%%%%%%%%%%%%%%
\subsubsection{The coefficient $\cualpha$}
\label{sec:cu-alpha}

Another variation of the Krippendorff's $\alpha$ coefficient is the so-called $\cualpha$ coefficient. As the previous variation, this is a coefficient that is computed for each semantic domain, say $S_{i}$ for some $1 \leq i \leq s$. Suppose that this semantic domain contains codes $C^{i}_1, \ldots, C^{i}_r$. As always, the set of considered items is the set of quotations. However, the collection of labels is now a set
$$
    \Lambda = \left\{\cC_1, \ldots, \cC_r\right\}.
$$
Semantically, they are labels that represent the codes of the chosen domain $S_{i}$.

For the assignment of labels to items, the rule is as follows. For $1 \leq \alpha \leq n$ and $1 \leq \beta \leq m$, we set $\omega_{\alpha,\beta} = \cC_k$ if the judge $J_\alpha$ assigned the code $C_k^{i}$ of $S_{i}$ to the item (quotation) $I_\beta$. Recall that, from the exclusion principle for codes within a semantic domain, the judge $J_\alpha$ applied at most one code from $S_{i}$ to $I_\beta$. If the judge $J_\alpha$ did not apply any code of $S_{i}$ to $I_\beta$, we set $\omega_{\alpha, \beta} = \emptyset$. From this set of judgements $\Omega_{cu}^{S_{i}} = \left\{\omega_{\alpha, \beta}\right\}$, $\cualpha$ is given as
$$
    \cualpha^{S_{i}} = \alpha(\Omega_{cu}^{S_{i}}).
$$

\begin{remark}
As explained in Remark \ref{ref:not-voted-item} of Appendix \ref{appendix-alpha}, for the computation of the observed and expected coincidence matrices, only items that received at least to evaluations with codes of $S_{i}$ from two different judges count. In particular, if a quotation is not evaluated by any judge (irrelevant matter), received evaluations for other domains but not for $S_{i}$ (matter that does not corresponds to the chosen domain) or only one judge assigned to it a code from $S_{i}$ (singled-voted), the quotation plays no role in $\cualpha$. This limitation might seem a bit cumbersome, but it could be explained by arguing that the presence/absence of $S_{i}$ is measured by $\alphabin^{S_{i}}$ so it will be redundant to take it into account for $\cualpha^{S_{i}}$ too.
\end{remark}

%%%%%%%%%%%%%%%%%%%%%%%%%%%%%%%%%%%%%%%%%%%%%%%%%%%%%%%%%%%%%%%%%%%%%
\subsubsection{The coefficient $\Cualpha$}

The last variation of Krippendorff's $\alpha$ coefficient that we consider in this study is the so-called $\Cualpha$ coefficient. In contrast with the previous coefficients, this is a global measure of the goodness of the partition into semantic domains. Suppose that our codebook determines semantic domains $S_1, \ldots, S_s$. As always, the set of considered items is the set of quotations, but the collection of labels is the set
$$
    \Lambda = \left\{\cS_1, \ldots, \cS_s\right\}.
$$
Semantically, they are labels representing the semantic codes of our codebook.

We assign labels to items as follows. Let $1 \leq \alpha \leq n$ and $1 \leq \beta \leq m$. Then, if $\sigma_{\alpha, \beta} = \left\{C_{j_1}^{i_1}, \ldots, C_{j_p}^{i_p}\right\}$, we set $\omega_{\alpha, \beta} = \left\{\cS_{i_1}, \ldots, \cS_{i_p}\right\}$. In other words, we label $I_\beta$ with the labels corresponding the semantic domains chosen by judge $J_\alpha$ for this item, independently of the particular code. Observe that this is the first case in which the final evaluation $\Omega$ might be multivaluated. From this set of judgements, $\Omega_{Cu} = \left\{\omega_{\alpha, \beta}\right\}$, $\Cualpha$ is given as
$$
    \Cualpha = \alpha(\Omega_{Cu}).
$$

In this way, $\Cualpha$ measures the degree of reliability in the decision of applying the different semantic domains, independently of the particular chosen code. Therefore, it is a global measure that quantifies the logical independence of the semantic domains and the ability of the judges of looking at the big picture of the matter, only from the point of view of semantic domains.

\section{Appendix. Universal Krippendorff's $\alpha$ coefficient}
\label{appendix-alpha}
In this appendix, we rephrase Krippendorff's $\alpha$ for a wide class of judgements. This gives rise to a universal Krippendorff's $\alpha$ coefficient formulated for assignments of `meta-codes' called `labels'. This formulation is very useful for the unified formulation of several variants of $\alpha$, as introduced in Appendix \ref{appendix-ica}. For an historical description of this coefficient, check \cite{Krippendorff:2018}.

In the context of Inter-Coder Agreement analysis, we are dealing with $n > 1$ different judges (also known as coders), denoted by $J_1, \ldots, J_n$, as well as with a collection of $m \geq 1$ items to be judged (also known as quotations), denoted $I_1, \ldots, I_m$. We fix a set of $N \geq 1$ admissible `meta-codes', called labels, say $\Lambda = \left\{ l_1,\ldots, l_N\right\}$. The task of each of the judges $J_\alpha$ is to assign, to each item $I_\beta$, a collection (maybe empty) of labels from $\Lambda$. Hence, as byproduct of the evaluation process, we get a set $\Omega = \left\{\omega_{\alpha, \beta}\right\}$, for $1 \leq \alpha \leq n$ and $1 \leq \beta \leq m$, where $\omega_{\alpha, \beta} \subseteq \Lambda$ is the set of labels that the judge $J_\alpha$ assigned to the item $I_\beta$. Recall that $\omega_{\alpha, \beta}$ is not a multiset, so every label appears in $\omega_{\alpha, \beta}$ at most once.

From the collection of responses $\Omega$, we can count the number of observed pairs of responses. For that, fix $1 \leq i,j \leq N$ and set
$$
    o_{i,j} = \left|\left\{(\omega_{\alpha, \beta}, \omega_{\alpha', \beta}) \in \Omega^2 \,\left|\, \alpha'\neq \alpha, \begin{tabular}{c}
         $\left(l_i \in \omega_{\alpha, \beta} \textrm{ and } l_j \in \omega_{\alpha', \beta} \right)$ \\ \textrm{or} \\ $\left(l_j \in \omega_{\alpha, \beta} \textrm{ and } l_i \in \omega_{\alpha', \beta}\right)$
    \end{tabular} \right.\right\}\right|.
$$
In other words, $o_{i,j}$ counts the number of (ordered) pairs of responses of the form $(\omega_{\alpha, \beta}, \omega_{\alpha', \beta}) \in \Omega \times \Omega$ that two different judges $J_\alpha$ and $J_{\alpha'}$ gave to the same item $I_\beta$ and such that $J_\alpha$ included $l_i$ in his response and $J_{\alpha'}$ included $l_j$ in his response, or viceversa.

\begin{remark}\label{ref:not-voted-item}
Suppose that there exists an item $I_\beta$ that was judged by a single judge, say $J_\alpha$. The other judges, $J_{\alpha'}$ for $\alpha' \neq \alpha$, did not vote it (or, in other words, they voted it as empty), so $\omega_{\alpha', \beta} = \emptyset$. Then, this item $I_\beta$ makes no contribution to the calculation of $o_{i,j}$ since there is no other judgement to which $\omega_{\alpha, \beta}$ can be paired. Hence, from the point of view of Krippendoff's $\alpha$, $I_\beta$ is not considered. This causes some strange behaviours in the coefficients below that may seem counterintuitive.
\end{remark}

From these counts, we construct the matrix of observed coincidences as $M_o = \left(o_{i,j}\right)_{i,j=1}^N$. By its very construction, $M_o$ is a symmetric matrix. From this matrix, we set $t_k = \sum_{j=1}^N o_{k, j}$, which is (twice) the total number of times that the label $l_k \in \Lambda$ was assigned by any judged. Observe that $t = \sum_{k=1}^N t_k$ is the total number of judgments. In the case that each judge evaluates each item with a single non-empty label, we have $t = nm$.

On the other hand, we can construct the matrix of expected coincidences, $M_e = \left(e_{i,j}\right)_{i,j = 1}^N$, where
$$
    e_{i,j} = \left\{\begin{array}{cc}
        \frac{t_i}{t}\frac{t_j}{t-1}t = \frac{t_it_j}{t-1} & \textrm{if }i\neq j \\
        & \\
         \frac{t_i}{t}\frac{t_i-1}{t-1}t = \frac{t_i(t_i-1)}{t-1} & \textrm{if }i = j
    \end{array}\right.
$$
The value of $e_{i,j}$ might be though as the average number of times that we expect to find a pair $(l_i, l_j)$, when the frequency of the label $l_i$ is estimated from the sample as $t_i/t$. Again, $M_e$ is a symmetric matrix.

Finally, let us fix a pseudo-metric $\delta: \Lambda \times \Lambda \to [0, \infty) \subseteq \mathbf{R}$, i.e.\ a symmetric function satisfying the triangle inequality and with $\delta(l_i,l_i)=0$ for any $l_i \in \Lambda$ (recall that this is only a pseudo-metric since different labels at distance zero are allowed). This metric is given by the semantic of the analyzed problem and, thus, it is part of the data used for quantifying the agreement. The value $\delta(l_i, l_j)$ should be seen as a measure of how similar the labels $l_i$ and $l_j$ are. A common choice is so-called discrete metric, given by $\delta(l_i, l_j) = 0$ if $i = j$ and $\delta(l_i, l_j) = 1$ otherwise. The discrete metric means that all the labels are equally separated. This is the underlying assumption that we will apply in our study. However, subtler metrics may be used for extracting more semantic information from the data (see \cite{Krippendorff:2016}).

From these computations, we define the observed disagreement, $D_o$, and the expected disagreement, $D_e$, as
\begin{equation}\label{eq:disagreement}
    D_o = \sum_{i=1}^N\sum_{j=1}^N o_{i,j} \delta(l_i, l_j), \hspace{0.5cm} D_e = \sum_{i=1}^N\sum_{j=1}^N e_{i,j}\delta(l_i, l_j).
\end{equation}
These quantities measure the degree of disagreement that is observed from $\Omega$ and the degree of disagreement that might be expected by judging randomly, respectively.

\begin{remark}
In the case of taking $\delta$ as the discrete metric, we have another interpretation of the disagreement. Observe that, in this case, since $\delta(l_i, l_i)=0$ we can write the disagreements as
$$
    D_o = \sum_{i\neq j} o_{i,j} = t - \sum_{i=1}^N o_{i,i}, \hspace{0.5cm} D_e = \sum_{i\neq j} e_{i,j} = t - \sum_{i=1}^N e_{i,i}.
$$
The quantity $A_o = \sum_{i=1}^N o_{i,i}$ (resp.\ $A_e = \sum_{i=1}^N e_{i,i}$) can be understood as the observed (resp.\ expected) agreement between the judges. In the same vein, $t = \sum_{i,j=1}^N o_{i,j} = \sum_{i,j=1}^N e_{i,j}$ may be seen as the maximum achievable agreement. Hence, in this context, the disagreement $D_o$ (resp.\ $D_e$) is indeed the difference between the maximum possible agreement and the observed (resp.\ expected) agreement.
\end{remark}

From these data, Krippendroff's $\alpha$ coefficient is defined as
$$
    \alpha = \alpha(\Omega) = 1 - \frac{D_o}{D_e}.
$$
From this formula, observe we are following limitting values:
\begin{itemize}
    \item $\alpha = 1$ is equivalent to $D_o = 0$ or, in other words, it means that there exists perfect agreement in the judgements among the judges.
    \item $\alpha = 0$ is equivalent to $D_o = D_e$, which means that the agreement observed between the judgements is entirely due to chance.
\end{itemize} 
In this way, Krippendorff's $\alpha$ can be interpreted as a measure of the degree of agreement that is achieved out of the chance. The bigger the $\alpha$ is, the better agreement is observed.

\begin{remark}
Observe that $\alpha < 0$ may only be achieved if $D_o > D_e$, which means that there is even more disagreement than the one that could be expected by chance. This implies that the judges are, consistently, issuing different judgements for the same items. Thus, it evidences that there exists an agreement between the judges to not agree, that is, to fake the evaluations. On the other hand, as long as the metric $\delta$ is non-negative, $D_o \geq 0$ and, thus, $\alpha \leq 1$.
\end{remark}

%\begin{acknowledgements}
%\end{acknowledgements}

\bibliography{references.bib}{}
\bibliographystyle{abbrv}

\end{document}